\newcommand{\CLASSINPUTbaselinestretch}{1.5}
\newcommand{\CLASSINPUTinnersidemargin}{1in}
\newcommand{\CLASSINPUTtoptextmargin}{1in}

\documentclass[journal,draftcls,onecolumn,12pt,twoside]{IEEEtranTCOM}
\usepackage{graphicx}
\usepackage{epstopdf}
\DeclareGraphicsExtensions{.eps, .pdf}

\usepackage{amsfonts}
\usepackage{rotating}
\usepackage[caption=false]{subfig}
\usepackage[cmex10]{amsmath} 
\usepackage{array}
\usepackage{cite}
\usepackage{amssymb}
\usepackage{pifont}
\usepackage{amsfonts}
\usepackage{amsmath}
\usepackage{stackrel}
\usepackage{booktabs,multirow}
\usepackage{arydshln}
\usepackage{slashbox}
\usepackage{amsthm}
\usepackage{listings}
\usepackage{algorithmicx}
\usepackage{algorithm}
\usepackage{algpseudocode}
\usepackage{tablefootnote}
\usepackage{threeparttable}
\usepackage{hhline}
\DeclareMathAlphabet{\mathbcal}{OMS}{cmsy}{b}{n}
\DeclareMathAlphabet{\mathcal}{OMS}{cmsy}{b}{n}
\DeclareMathAlphabet{\mathfrak}{OMS}{cmsy}{b}{n}

\newtheorem{rem}{Remark}

\newtheorem{ex}{Example}

\newfloat{routine}{htbp}{loa}
\floatname{routine}{Routine} 
\makeatletter
\newcommand{\algmargin}{\the\ALG@thistlm}
\makeatother
\newlength{\forwidth}
\algdef{SE}[parFOR]{parFor}{EndparFor}[1]
  {\parbox[t]{\dimexpr\linewidth-\algmargin}{
     \hangindent\forwidth\strut\algorithmicfor\ #1\ \algorithmicdo\strut}}{\algorithmicend\ \algorithmicfor}
\algnewcommand{\parState}[1]{\State
  \parbox[t]{\dimexpr\linewidth-\algmargin}{\strut #1\strut}}

\newlength{\ifwidth}
\algdef{SE}[parIF]{parIf}{EndparIf}[1]
  {\parbox[t]{\dimexpr\linewidth-\algmargin}{
     \hangindent\ifwidth\strut\algorithmicif\ #1\ \algorithmicdo\strut}}{\algorithmicend\ \algorithmicif}

\graphicspath {{figures/}}

\begin{document}
\title{Error Floor Analysis of LDPC Column Layered
Decoders}

\author{Ali Farsiabi, and Amir H. Banihashemi}

%

\maketitle

\begin{abstract}
In this paper, we analyze the error floor of column layered decoders, also known as shuffled decoders, for low-density parity-check (LDPC) codes
under saturating sum-product algorithm (SPA). 
To estimate the error floor, we evaluate the failure rate of different trapping sets (TSs) that contribute to the frame error rate in 
the error floor region. For each such TS, we model the dynamics of SPA in the vicinity of the TS by a linear state-space model that incorporates the information of the 
layered message-passing schedule. Consequently, the model parameters and the failure rate of the TS change as a result of the change in the order by which the 
messages of different layers are updated. This, in turn, causes the error floor of the code to change as a function of scheduling. Based on the proposed analysis, we then 
devise an efficient search algorithm to find a schedule that minimizes the error floor. Simulation results are presented to verify the accuracy of the proposed error floor estimation technique.
\end{abstract}


\IEEEpeerreviewmaketitle
\section{Introduction}
\IEEEPARstart {F}{inite}-length low-density parity-check (LDPC) codes suffer from an \emph{error floor} problem, i.e., as the quality of the channel improves, at a certain point, the improvement in the error rate performance does not follow its initial rapid rate. The error floor is attributed to certain harmful graphical structures within the code's Tanner graph, referred to as {\em trapping sets (TSs)}.
The error floor of LDPC codes has been widely studied from different perspectives including code constructions with low error floor~\cite{Ivkovic, Asvadi, Khaz, Nguyen, mao2001heuristic,xiao2004improved,Sima-CL1,Sima-CL2,Sima-TCOM,Bashir-TCOM,Bashir-TCOM2,Tian-2004,Peg,zheng2010constructing,Tao-2018}, decoder design for improved error floor performance~\cite{danesh, Ryan2,Ryan1, Kyung, Zhang1,homayoon2020,TB,TSbreaking,zhang_quasiuniform}, enumeration of TS structures~\cite{yoones2015, hashemireg, hashemiireg, mehdi2014, mehdi2012,Wang}, and error floor estimation~\cite{richardson,Cole, LaraIS, daneshrad,Lara_SP,Xiao,Sun_phd, Schleg, But_SS, Homayoon_SP,Sina,Ontology,Hu_magneticIS,XB-2007,Ali-TCOM}.


The analysis and estimation of the error floor of LDPC codes have generally been performed within two main categories of methods. 
In the first category, referred to as \emph{code-dependent}, the harmfulness of TSs is studied within the Tanner graph of a specific code. 
In other words, within this category of techniques, the information of the entire Tanner graph is required for the analysis. 
The majority of code-dependent methods are based on \emph{importance sampling} techniques~\cite{richardson,Cole, LaraIS, daneshrad,Lara_SP,Xiao,Hu_magneticIS,XB-2007}.  In the second category, known as \emph{code-independent}, 
the harmfulness of TSs are studied in isolation, and one would only need the topology of the TS and the degree distributions 
of the code (Tanner graph), rather than the entire 
Tanner graph, to estimate the error floor~\cite{Sun_phd, Schleg, But_SS,AS_threshold,Homayoon_SP,Ontology,Ali-TCOM}. 
Of particular interest in this work is the linear state-space model, proposed by Sun~\cite{Sun_phd}, and then modified and extended in \cite{But_SS,Schleg},
to estimate the error floor of LDPC codes decoded by sum product algorithm (SPA) over the additive white Gaussian noise (AWGN) channel.
This model estimates the behavior of SPA in the vicinity of a TS by a linear state-space model, in which the state variables are the messages inside the TS and
the behavior of messages outside the TS is approximated by density evolution (DE)~\cite{Urbank}.

The iterative message-passing algorithms, commonly used to decode LDPC codes, can be applied based on various schedules~\cite{L1,L2,XiaoSchedule,NouhSchedule,L4,L9
,L6,L8,L3,L5,L7,CL1,CL2,CL3,CL4,CL5,CL6,CL7}
For example, in \emph{two-phase message passing} or \emph{flooding schedule}, each iteration is divided into two phases. 
In the first (second) phase, all the variable nodes (VNs) (check nodes (CNs)) of the code's Tanner graph pass their messages 
towards their adjacent CNs (VNs). There are also various types of \emph{layered} or \emph{serial} message-passing schedules that 
may be favorable to flooding schedule due to lower complexity and higher convergence speed~\cite{L1,L2,L3,L4,L5,L6,L7,L8,L9,XiaoSchedule,NouhSchedule,InfromedDynamic_wesel_2010,M2I2}. In particular, in column (row) layered decoders, 
the VNs (CNs) of the Tanner graph are partitioned into different groups known as {\em layers}. Correspondingly, each iteration is divided into different sub-iterations that are performed sequentially across the layers. Within each sub-iteration, only the messages corresponding to a specific layer are updated while other layers remain inactive. As a result, the updated reliabilities of each layer can be utilized within the subsequent layers, leading to a higher convergence speed compared to flooding schedule. 
The layered decoders are often used along with quasi-cyclic (QC) LDPC codes. In this case, the row or column blocks of the parity-check matrix of the QC-LDPC code are 
treated as layers. In layered decoders for QC-LDPC codes, since one can reuse the same hardware resources for different sub-iterations, the implementation complexity 
can be significantly reduced compared to decoders with flooding schedule.  

The majority of existing works on theoretical analysis of error floor in LDPC codes, including the linear state-space model of~\cite{Sun_phd, Schleg, But_SS}, are based on flooding schedule. With the exception of~\cite{Ali-ITrow}, the error floor study of column (or row) layered decoders however, 
has been mainly limited to empirical results~\cite{InfromedDynamic_wesel_2010,M2I2,Angarita_MS_2014,BackTrack_hard_2011,
IDS_kim_2012let,vasic_horizontal}. This is despite the fact that layered decoders are widely used in practical applications. 
In~\cite{InfromedDynamic_wesel_2010} and \cite{M2I2}, dynamic scheduling and schedule diversity, respectively, have been shown to lower the error floor. 
More recently, Raveendran and Vasic \cite{vasic_horizontal} have demonstrated that the error floor of $(155,64)$ Tanner code 
decoded by row layered Gallager-B algorithm over the binary symmetric channel (BSC) is lower than 
the case where the decoding is performed with the flooding schedule.

Most recently, in \cite{Ali-ITrow}, the error floor of row layered SPA over the AWGN channel was theoretically analyzed. 
The analysis was performed by developing a linear state-space model for TSs that incorporates the layered nature of the decoding algorithm. 
The analysis in \cite{Ali-ITrow} shows that the harmfulness of a given TS, as well as the error floor of the code, can significantly change by 
changing the order in which the information of different layers, corresponding to different row blocks of the parity-check matrix, are updated. 
The proposed model  is then used in \cite{Ali-ITrow} to optimize the error floor performance of the row layered decoders.

In this paper, we extend the linear state-space model of \cite{Ali-ITrow} to column layered decoders, also known as {\em shuffled} decoders~\cite{L7,L8,CL1,CL2,CL3,CL4,CL5,CL6,CL7}. Undertaking such an analysis is important as column layered decoders are often implemented in practical applications, particularly those involving high-rate LDPC codes. The extension of the linear model of~\cite{Ali-ITrow} to column layered decoders involves a number of non-trivial 
steps including the derivation of the model parameters. 
In addition to developing the model to analyze the failure rate of a TS, we also use the proposed model to optimize the error floor performance 
of column layered decoders. As the number of column layers is larger than the number of row layers for a given QC-LDPC code, the size of the search space for such an optimization is larger than that of a similar optimization for row layered decoders. The difference in size of the search space 
is especially large for high-rate codes. This makes the corresponding optimization more difficult to carry out.

Finally, we compare our theoretical estimates of the error floor with Monte Carlo simulation results for QC-LDPC codes, 
both variable-regular and irregular, and demonstrate a good match between the two. The results show that the error floor of a given code decoded by a column layered decoder can in general be better or worse than that of a row layered decoder, depending on the choice of the scheduling. We also observe that, for the tested codes, 
the performance variation among different schedules is larger for row layered decoders compared to their column layered counterparts.

The rest of the paper is organized as follows: In Section~\ref{premSecCol}, we present some preliminaries. This is followed by a review of  the linear state-space model of a 
TS for a decoder with the layered schedule in Section~\ref{SS_Col_Sec}. Within the same section, we obtain the model parameters related to column layered decoders. In Section \ref{col_opt_Sec}, we discuss our approach to optimize the updating order of the layers in a column layered decoder for minimizing the error floor. Section~\ref{SimulaColSection} is used to present the simulation results and evaluate the accuracy of the developed theoretical analysis in estimating the error floor.

\section{Preliminaries: LDPC codes, SPA decoding, column layered scheduling and TSs} \label{premSecCol}

In this work, we consider the application of a binary LDPC code in the transmission of binary data over an AWGN channel.
For modulation, we assume binary phase shift keying (BPSK) at the output of LDPC encoder. The LDPC code is defined 
as the null space of a sparse parity-check matrix $\mathbf{H}$, i.e., any codeword $\bold{d}$ satisfies  $\mathbf{H}\bold{d} = \bold{0}$.
At the BPSK modulator, codeword bits $0$ and $1$ are mapped to the constellation points $+1$ and $-1$, respectively.
At the time index $i$, the channel output $y_i$ is given by $y_i=u_i+n_i$, where $u_i \in \{+1,-1\}$, and $n_i$ is 
a zero-mean Gaussian random variable with variance ${\sigma_{ch}^2}$. 


Any LDPC code can be represented by a bipartite graph $G=(V \cup C,E)$, called a {\em Tanner graph}. The two sets of nodes $V=\{ v_1,v_2,\dots,v_n \}$ and 
$C=\{ c_1,c_2,\dots,c_m \}$ are referred to as {\em variable nodes} (VNs) and {\em check nodes } (CNs), respectively. The set of edges $E=\{ e_1,e_2,\dots,e_k \}$
represents the connections between VNs and CNs. A Tanner graph $G$ corresponds to a parity-check matrix $\mathbf{H}$ of the code, i.e., a VN, $v_i$, is connected to a CN, $c_j$, via an edge in $G$ if and only if the corresponding element in $\mathbf{H}$ is one, $\mathbf{H}(j,i)=1$.
The parameter $n$, i.e., the number of variable nodes, is the code's block length, and the code rate ${\cal R}$ satisfies ${\cal R} \geq 1 - m/n$.



In our model, when a noisy version of a codeword is received at the receiver, we use SPA to recover the transmitted codeword.
We assume that the input of the SPA is the log-likelihood ratio (LLR) of the codeword bits. Let ${v_i}$ represent the $i$th VN corresponding to the $i$th symbol of the received block. The LLR of this VN over the binary AWGN channel is given by
\begin{small}
\begin{equation}
{L^{ch}_i=2y_i/\sigma_{ch}^2}.
\label{CH_LLR}
\end{equation}
\end{small}
We use the notation ${M(i)}$ (${N(j)}$) to represent the set of CNs (VNs) adjacent to VN ${v_i}$ (CN $c_j$). 
To denote the message sent from VN ${v_i}$ (CN ${c_j}$) to CN ${c_j}$ (VN ${v_i}$) at iteration $\ell$, we use the notation ${L_{\ell}^{[i \rightarrow j]}}$
(${L_{\ell}^{[i \leftarrow j]}}$). In the SPA, the messages sent from VNs to CNs at iteration $\ell$ are given by
\begin{small}
\begin{equation}
{{L_{\ell}^{[i \rightarrow j]}}=L^{ch}_i+\sum_{k \in M(i)\backslash j}{L_{\ell-1}^{[i \leftarrow k]}}}\:,
\label{v2c_message}
\end{equation}
\end{small}
in which $ M(i)\backslash j$ is used for all CNs adjacent to VN $v_i$ excluding CN $c_j$.
At the first iteration of the algorithm, the VN to CN messages in (\ref{v2c_message}) are initialized by the channel LLRs given in (\ref{CH_LLR}).
The CN to VN messages in SPA are computed as
\begin{small}
\begin{equation}
{L_{\ell}^{[i \leftarrow j]}=2\tanh^{-1}\Bigg[{\prod_{k \in N(j)\backslash i }\tanh\frac{{L_{\ell}^{[k \rightarrow j]}}}{2}}\Bigg]}.
\label{SPA_CN}
\end{equation} 
\end{small}
At the end of each iteration $\ell$, first, the total LLR, 
\begin{small}
\begin{equation}
  {{\tilde{L}_{\ell}^{[i]}}=L^{ch}_i+\sum_{k \in M(i)}{L_{\ell}^{[i \leftarrow k]}}},
  \label{TotLLR}
\end{equation}
\end{small}
is calculated for each coded bit, and then, a hard decision is made 
\begin{small}
\begin{equation}
\hat{d_i}=[\text{sign}({\tilde{L}_{\ell}^{[i]}})+1]/2.
\end{equation}
\end{small}
At the end of the ${\ell}$-th iteration, if $\mathbf{H}\hat{\bold{d}}=\bold{0}$ is satisfied for the decoded block $\hat{\bold{d}}$, then
the decoding operation is terminated. Otherwise, the iterative decoding continues until the maximum number of iterations, $I_{max}$, is reached. 
At that point, if there still exists any unsatisfied parity-check equation, a decoding failure is declared.

In this work, similar to~\cite{butler_numerical,zhang_quasiuniform}, we use the following reformulation of (\ref{SPA_CN}), 
which is more robust against numerical errors:
\begin{small}
\begin{equation}
{L_{\ell}^{[i \leftarrow j]}=\underset{k \in N(j)\backslash i }{\boxplus}}{L_{\ell}^{[k \rightarrow j]}}\:.
\label{boxplus1}
\end{equation}
\end{small}
In (\ref{boxplus1}), the pairwise box-plus operator, $\boxplus$, is defined as
\begin{small}
\begin{equation}
\begin{split}
x_1 \boxplus x_2 & = \ln\Bigg(\dfrac{1+e^{x_1+x_2}}{e^{x_1}+e^{x_2}}\Bigg)    =\text{sign}(x_1)\text{sign}(x_2).\min(|x_1|,|x_2|)+s(x_1,x_2)\:,
\end{split}
\label{boxplus_pair}
\end{equation}
\end{small}
where the term $s(x_1,x_2)$ is given by
\begin{small}
\begin{equation}
{s(x_1,x_2)=\ln\Bigg(1+e^{-|x_1+x_2|}
\Bigg)-\ln\Bigg(1+e^{-|x_1-x_2|}
\Bigg).}
\label{s_correction}
\end{equation}
\end{small}
It should be noted that the equations \eqref{v2c_message}-\eqref{boxplus1} are edgewise operations, and thus they can be executed with various message-passing schedules. In \emph{flooding schedule}, at the first half of each iteration, equation \eqref{v2c_message} is executed for all the VN to CN messages 
and in the second half, all the CN to VN messages are updated based on \eqref{SPA_CN}. Then, using \eqref{TotLLR}, the LLRs of all VNs are updated simultaneously. 
In \emph{column layered schedule}, however, the VNs are divided into a number of subgroups (layers), and the messages within an iteration are 
sequentially updated, one layer at a time, and based on the order in which the layers are scheduled to be updated. For each layer, 
based on the incoming messages from the adjacent CNs, the total LLRs of the corresponding VNs of 
the current layer are updated. The incoming messages from the adjacent CNs, themselves,
are calculated based on the extrinsic LLRs received by such CNs from the VNs of other layers. Depending on the updating order of layers, these LLR messages  
may have been updated in the previous iteration or in the layers of the current iteration that have been updated earlier.
When the LLR values of the VNs within a layer are all updated, these updated values will be used in updating the messages of the subsequent layers. 

In Algorithm \ref{ColLayered_alg}, the steps of the column layered SPA for QC-LDPC codes are presented. In this algorithm,
it is assumed that the Tanner graph of a QC-LDPC code is constructed based on a \emph{cyclic p-lifting} of a smaller 
Tanner graph, known as \emph{base graph}~\cite{MKarimi_girth}. The $m_b \times n_b$ bi-adjacency matrix, $H_b$, of the base graph is referred to as \emph{base matrix}. 
As a result of cyclic lifting, the  parity-check matrix of the QC-LDPC code consists of an $m_b \times n_b$ array of $p \times p$ circulants or all-zero matrices,
depending on whether the corresponding element in $H_b$ is one or zero, respectively.
The cyclic lifting thus imposes a structure on the Tanner graph of QC-LDPC codes such that the VNs and CNs can be grouped into 
$n_b$ and $m_b$ different types, respectively, where each type corresponds to one row or column block of the parity-check matrix. 
Type-$i$ VNs are connected to Type-$j$ CNs by a cluster of edges if the corresponding element at column $i$ and row $j$ of $H_b$ is $1$.   

\begin{algorithm}
\small
\caption{The Box-Plus Column Layered Decoding Algorithm }
\label{ColLayered_alg}

 \begin{algorithmic} [1] 
 \State \textbf{Input:} Channel LLRs for all the VNs. 
 \State {\textbf{Initialization:} All the VN to CN messages are initialized with channel LLRs, $L^{ch}_i$, for $i=1,\dots,n_b \times p$. }
 \For{iteration $\ell=1,\dots,I_{max}$}
 \For{column block $z=1,\dots,n_b$}
 \For{VNs $i=(z-1)p+1,\dots,zp$, in layer $z$} 
 \ForAll{CNs $j \in M(i)$}
 \State{${L^{[i \leftarrow j]}=\underset{k' \in N(j)\backslash i }{\boxplus}}{L^{[k' \rightarrow j]}}$,}
 \EndFor
 \State{${{L^{[i \rightarrow j]}}=L^{ch}_i+\sum_{k \in M(i)\backslash j}{L^{[i \leftarrow k]}}}$.}
 \State{${\tilde{L}^{[i]}}=L^{ch}_i+\sum_{k \in M(i)}{L^{[i \leftarrow k]}}$.}
 \EndFor
  \EndFor
 \par{Hard-decision:}
 \ForAll{VNs $i=1,\dots,n_b \times p$}
 \State{$\hat{d_i}=[\text{sign}({\tilde{L}^{[i]}})+1]/2$.}
 \EndFor
 \If{$\bold{H}\hat{\bold{d}}=\bold{0}$}
 \State{Break.}
 \EndIf
 \EndFor 
\State \textbf{Output: $\hat{\bold{d}}$} 
 \end{algorithmic}
 \end{algorithm}


Let $\mathcal{S}$ be a subset of VNs, $V$, and $\Gamma{(\mathcal{S})}$ be the subset of CNs $C$ that 
are adjacent to $\mathcal{S}$ in the Tanner graph $G=(V \cup C, E)$. 
The \textit{induced subgraph} of $\mathcal{S}$ in $G$, denoted by $G(\mathcal{S})$, is a graph whose nodes and edges are 
$\mathcal{S} \cup \Gamma{(\mathcal{S})}$ and $\{v_i c_j \in E : v_i \in \mathcal{S}, c_j \in \Gamma{(\mathcal{S})}\}$, respectively. The set of CNs, $\Gamma{(\mathcal{S})}$, can be partitioned into even-degree CNs, $\Gamma_{e}{(\mathcal{S})}$, and odd-degree CNs, $\Gamma_{o}{(\mathcal{S})}$. The members of $\Gamma_{o}{(\mathcal{S})}$ and $\Gamma_{e}{(\mathcal{S})}$ are referred to as \textit{unsatisfied check nodes} and \textit{satisfied check nodes}, respectively.  

An \textit{(a,b) trapping set (TS)} is a set $\mathcal{S}\subset V$, with $|\mathcal{S}|=a$ and $|\Gamma_{o}{(\mathcal{S})}| = b$. 
It is, alternatively, said that $\mathcal{S}$ belongs to the {\em class of (a,b) TSs}. When a TS $\mathcal{S}$ is in error in the error floor region, 
it often happens that the rest of variable nodes in $V$ are decoded correctly. Under such circumstances, all the check equations corresponding to the CNs in $\Gamma_{e}{(\mathcal{S})}$ are satisfied, and those that correspond to $\Gamma_{o}{(\mathcal{S})}$ are unsatisfied. Since the bits in $\mathcal{S}$ are in fact in error when the decoder is trapped in $\mathcal{S}$, we refer to CNs of $\Gamma_{e}{(\mathcal{S})}$ as being \textit{mis-satisfied}. In this paper, we sometimes refer to 
the subgraph $G(\mathcal{S})$, rather than the set of VNs $\mathcal{S}$, as the TS. This will be clear from the context.
\vspace{-0.944cm}

If all the CNs of $G(\mathcal{S})$ have degrees $1$ or $2$, the TS is called an \textit{elementary trapping set (ETS)}. An ETS 
is referred to as being {\em leafless (LETS)} 
if all of its VNs are connected to at least $2$ mis-satisfied CNs. Our experiments show that the vast majority of the structures that trap the SPA on a QC-LDPC code over the AWGN channel are LETSs. In this work therefore, similar to \cite{But_SS,Schleg,Ali-ITrow}, we focus on LETSs for our study. 

\section{State Space Model of Column Layered Decoders in Error Floor Region} \label{SS_Col_Sec}

\subsection{State-Space Model} \label{SS_Col_subSec}
It was shown in~\cite{But_SS,Schleg,Sun_phd} that the dynamics of a message-passing algorithm with flooding schedule in the vicinity of a TS can be well approximated 
by a linear state-space model in the error floor region. 
A similar model for row layered decoders was developed in~\cite{Ali-ITrow}. 
%
In this work, we extend the state-space model of \cite{Ali-ITrow} to column layered decoders. While the general formulation of state-space equations is the same for the two cases, the calculation of the model parameters differ significantly.


The state-space model of a LETS for a layered decoder is given by~\cite{Ali-ITrow}:
\begin{small}
\begin{IEEEeqnarray*}{lCl"s}
\tilde{\mathbf{x}}^{(0,j)}=\bold{0}, \ \ \ \ \ \ \ \  \ \ \ \ \ \ \  \:\:\:\:\:\:\:\:\:\:\:\: \:\:\:\:\:\:\:\:\:\:\: \:\:\:\:\: \:\:\:\:\:\:\: \:\:\:\:\: \:\:\:\: \: \: \:\:\:\:\ \ \ \ \ \ \ \ \ \  \ \ \ \ \ \text{\ \ \ \  \ \ \ \ \ \ for \ \ $\ell=0$, $ 1 \leq j \leq J$\:,} \IEEEyesnumber\\
\label{lay_sj}
\tilde{\mathbf{x}}^{(\ell,j)}= \  \mathfrak{G}_j^{(\ell)}\big (\mathfrak{A}_j\tilde{\mathbf{x}}^{(\ell-\delta_{j1}, j-1+J\delta_{j1})} +\mathfrak{B}_j\L+\overset{\triangleleft}{\mathfrak{B}}_{exj}\L_{ex}^{(\ell-1)}+\overset{\triangleright}{\mathfrak{B}}_{exj}\L_{ex}^{(\ell)} \big ), \:\:\:\text{for \ \ $\ell\geq 1$, $ 1 \leq j \leq J$.\IEEEyesnumber}
\end{IEEEeqnarray*} 
\end{small}
In the above equations, the vector $\tilde{\mathbf{x}}^{(\ell,j)}$ is of size $m_s$ and contains the state variables (messages passed over different edges of the LETS subgraph) 
at layer $j$ of iteration $\ell$. The vectors $\L$, $\L_{ex}^{(\ell-1)} $ and $\L_{ex}^{(\ell)} $ are the inputs of the model from the channel, and from unsatisfied CNs at iterations $\ell-1$ and $\ell$, respectively. We note that $\L_{ex}^{(0)}=\bold{0}$. The symbol $J$ denotes the total number of layers involved in the LETS subgraph.
While for a row layered decoder, this is determined by the number of layers in which the missatisfied CNs of the structure are involved, for a column layered decoder, this  parameter is determined by the number of layers to which the VNs of the structure belong. 
The updating rules for the state variables that are activated in layer $j$ are governed by the $m_s \times m_s$ transition matrix of this layer, denoted by 
$\mathcal{A}_j$. This matrix is equal to an $m_s \times m_s$ identity matrix in which the rows that correspond to the state variables of the $j$th layer 
are replaced with vectors that represent the message-passing mechanism of layer $j$ (how the state variables of the $j$th layer are a function of other state variables). 
Matrix $\mathcal{B}_j$ is an $m_s \times a$ matrix that indicates the contribution of channel LLRs in the calculation of the state variables of the $j$th layer. Similarly,  matrices $\overset{\triangleleft}{\mathfrak{B}}_{exj}$ and $\overset{\triangleright}{\mathfrak{B}}_{exj}$ are defined to account for the contribution of $\L_{ex}^{(\ell-1)}$ and $\L_{ex}^{(\ell)}$ in updating the state variables within the $j$-th layer, respectively. Also, $\delta_{j1}$ is the Kronecker delta function which is equal to $1$ when $j=1$ and is zero, otherwise. Equation (\ref{lay_sj}) implies that at the first layer of every iteration, the state vector is updated based on the last layer of the previous iteration while in the other layers, the updated states are a function of the state vector at the previous layer within the same iteration. The effect of mis-satisfied CNs within a given layer $j$ at iteration $\ell$ is modeled by the iteration dependent gain matrix $\mathfrak{G}_j^{(\ell)}$. This is an $m_s \times m_s$ identity matrix whose diagonal elements 
that correspond to the state variables of layer $j$ are replaced with the average gains obtained from the linear approximation of the mis-satisfied CN operation impacting those state variables. 

\begin{figure}
\centering
\includegraphics[width=1.8in]{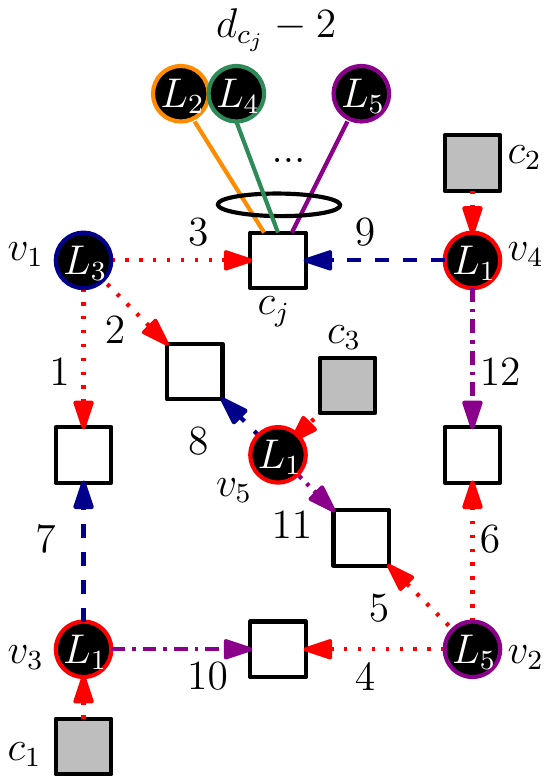}
\vspace{-0.4cm}
\caption{A $(5,3)$ LETS of the Tanner $(155, 64)$ code, in which all the state variables (internal messages) are labeled. The VNs, mis-satisfied CNs and unsatisfied CNs are shown by black circles, white squares and gray squares, respectively. The messages related to the layer numbers 1, 2 and 3 are represented by red (dotted line), blue (dashed line) and purple (dash-dotted line), respectively.}
\label{(5,3)collay_col}
\end{figure}
In the model presented here, when a layer is updated, the VNs of the layer 
receive messages from their adjacent CNs. These messages are computed at the CNs based on the messages received from the VNs of other layers. 
The number of VN layers in an LETS is generally less than or equal to the total number of column layers of the code, which is the same as the number of column blocks of the 
parity-check matrix, i.e., $J \leq n_b$. We use the notation $L_i~(i=1,\dots,n_b)$ to denote the $i$-th column layer of the code. The column layers are updated in the increasing order of their indices, i.e., layer $L_1$ is updated first, followed by $L_2$ and so on. Similarly, the layers of a TS are denoted 
by $L'_j~(j=1,\dots,J)$. The TS layers $L'_j$ are a subset of the code layers $L_i$.   

Following the extrinsic message-passing mechanism in column layered decoders, at each layer, certain internal messages of the TS incoming to some mis-satisfied CNs 
are activated. These mis-satisfied CNs are those adjacent to the VNs of the current layer, and the activated messages are those from the VNs of other layers that are adjacent to the said mis-satisfied CNs. In the following, we refer to these activated messages as the state variables of the corresponding layer. 
It is noted that, in column layered decoders, the two incoming messages to a specific mis-satisfied CN 
are activated at two different layers. This is one of the main differences between column layered decoders and row layered decoders, in which the incoming messages of a mis-satisfied CN are activated in the same layer.
 This is further explained in the following example.
 \vspace{-2cm} 
\begin{ex}
The QS Tanner $(155, 64)$ code has $5$ column layers, i.e., $n_b=5$. The VNs involved in the $(5,3)$ LETS of this code, shown in Fig. \ref{(5,3)collay_col}, belong to 3 different layers, i.e., $J=3$. These layers are $L'_1=L_1$, $L'_2=L_3$ and $L'_3=L_5$. For each of the three layers, the corresponding VNs and the internal messages  that are activated in that layer are shown by different colors and line types. For example, all the dotted red arrows are the messages that are passed from VNs to mis-satisfied CNs in the first layer of the column layered decoder. 
\end{ex}
\vspace{-2cm}
Suppose that in a LETS, the number of messages (state-variables) in layer $L'_j, j=1,\dots,J$, is denoted by $n_{L'_j}, j=1,\dots,J$. Since the total number of state variables is $m_s$, we have $n_{L'_1}+n_{L'_2}+\dots+n_{L'_J}=m_s$. Assume that the state variables of different layers are indexed with consecutive numbers, i.e., the state variables of the first layer are labeled as $x_1, \dots, x_{n_{L'_1}}$, the state variables of the second layer are labeled as $x_{n_{L'_1}+1}, \dots, x_{n_{L'_1}+n_{L'_2}}$, and so on. This type of labeling, referred to as \emph{systematic labeling}, allows us to establish a simple relationship between 
the transition matrix of the flooding schedule, denoted by $\mathbf{A}$, and that of layer $j$ of layered schedule, $\mathcal{A}_j$. 
In fact, with systematic labeling, matrix $\mathbf{A}$ is a $J \times J$ array of matrices $\mathbf{A}_{ij}$, $1\leq i,j \leq J$, where
all the diagonal blocks are zero matrices, i.e., $\mathbf{A}_{i,i}=0$, for $i=1,\dots,J$, and the off-diagonal blocks $\mathbf{A}_{i,j}, i \neq j$,
have size $n_{L'_i}\times n_{L'_j}$.
Matrix $\mathcal{A}_j$ has the same block structure as matrix $\mathbf{A}$. In fact, the $j$th row block of the two matrices are the same, 
but the other row blocks of $\mathcal{A}_j$ are zero matrices, except the diagonal blocks that are identity matrices. 
Matrices $\mathcal{B}_j$, $\overset{\triangleleft}{\mathfrak{B}}_{exj}$ and $\overset{\triangleright}{\mathfrak{B}}_{exj}$, based on systematic labeling, can also be partitioned into $J$ row blocks. The $j$-th row block of these matrices relate the channel inputs, the inputs of unsatisfied CNs from the previous iteration and those inputs from the current iteration to the state variables of layer $j$, respectively. All the other row blocks of these matrices are equal to zero.

\begin{ex}
The state variables in the LETS structure of  Fig. \ref{(5,3)collay_col} are labeled systematically.
As a result, we have the following systematic form for $\mathbf{A}$:\vspace{-.5cm}

\begin{small}
\begin{IEEEeqnarray*}{lCl"s}
\mathbf{A}=
\left[
\begin{array}{c|c|c}
\mathbf{0} &\mathbf{A}_{12}&\mathbf{A}_{13}\\
\hline
\mathbf{A}_{21}& \mathbf{0}& \mathbf{A}_{23}\\
\hline
\mathbf{A}_{31}& \mathbf{A}_{32}&\mathbf{0}
\end{array}
\right]
=\left[
\begin{array}{c c c c c c| c c c| c c c}
0& 0 &0 &0 &0& 0 &0& 1& 1 &0& 0 &0 \\
0& 0& 0 &0& 0&0 &1& 0& 1& 0& 0& 0\\
0& 0& 0 &0& 0& 0 &1 &1 &0 &0 &0& 0\\
0 &0& 0& 0& 0& 0 &0& 0& 0 &0& 1& 1\\
0 &0& 0& 0 &0 &0& 0 &0 &0&1 &0& 1\\
0 &0 &0& 0& 0& 0 &0 &0& 0 &1 &1 &0\\
\hline
0 &0& 0& 1 &0 &0& 0 &0 &0&0 &0& 0\\
0 &0 &0& 0& 1& 0 &0 &0& 0 &0 &0 &0\\
0& 0 &0 &0& 0& 1 &0 &0 &0& 0 &0 &0\\
\hline
1& 0 &0 &0& 0& 0 &0& 0& 0& 0& 0& 0\\
0 &1& 0 &0&0& 0& 0& 0 &0 &0 &0 &0\\
0& 0& 1& 0& 0 &0 &0 &0 &0& 0& 0& 0\\
\end{array}
\right].
\end{IEEEeqnarray*}
\end{small}
Also, the transition matrix of the 2nd layer of the column layered schedule is given by
\begin{small}
\begin{IEEEeqnarray*}{lCl"s}
 \mathbcal{A}_2=
\left[
\begin{array}{c|c|c}
\mathbf{I} &\mathbf{0}&\mathbf{0}\\
\hline
\mathbf{A}_{21}& \mathbf{0}& \mathbf{A}_{23}\\
\hline
\mathbf{0}& \mathbf{0}&\mathbf{I}
\end{array}
\right],
\end{IEEEeqnarray*}
\end{small}
in which the notation $\mathbf{I}$ is used to represent an identity matrix. 
Matrices $\overset{\triangleleft}{\mathfrak{B}}_{ex2}$ and $\overset{\triangleright}{\mathfrak{B}}_{ex2}$, that determine the contribution of unsatisfied CN inputs at iterations $\ell-1$ and $\ell$ to the second layer state variables at iteration $\ell$, are \vspace{-0.5cm}
\begin{small}
 
\begin{IEEEeqnarray*}{lCl"s}
\overset{\triangleleft}{\mathfrak{B}}_{ex2}=
\left[
\begin{array}{c c c}
0& 0 &0  \\
0& 0 &0 \\
0& 0 &0 \\
0& 0 &0 \\
0& 0 &0 \\
0& 0 &0 \\
\hline
0& 0 &0 \\
0& 0 &0 \\
0& 0 &0 \\
\hline
0& 0 &0 \\
0& 0 &0 \\
0& 0 &0 \\
\end{array}
\right], \
\overset{\triangleright}{\mathfrak{B}}_{ex2}=
\left[
\begin{array}{c c c}
0& 0 &0  \\
0& 0 &0 \\
0& 0 &0 \\
0& 0 &0 \\
0& 0 &0 \\
0& 0 &0 \\
\hline
1& 0 &0 \\
0& 0 &1 \\
0& 1 &0 \\
\hline
0& 0 &0 \\
0& 0 &0 \\
0& 0 &0 \\
\end{array}
\right].
\end{IEEEeqnarray*} 
\end{small}
As can be seen, since all the inputs of unsatisfied CNs are updated at layer $L'_1=L_1$, all the elements of $\overset{\triangleleft}{\mathfrak{B}}_{ex2}$, corresponding to layer $L'_2=L_3$, are equal to zero and all the state variables of the second layer use the newly updated external inputs of unsatisfied CNs, $\L_{ex}^{(\ell)}$. 
Similarly, for $\mathcal{B}_2$, we have\vspace{-.5cm}
\begin{small}
\begin{IEEEeqnarray*}{lCl"s}
{\mathfrak{B}}_{2}=
\left[
\begin{array}{c c c c c}
0& 0 &0& 0 &0  \\
0& 0 &0& 0 &0 \\
0& 0 &0 & 0 &0\\
0& 0 &0 & 0 &0\\
0& 0 &0 & 0 &0\\
0& 0 &0 & 0 &0\\
\hline
0& 0 &1 & 0 &0\\
0& 0 &0 & 0 &1\\
0& 0 &0 & 1 &0\\
\hline
0& 0 &0 & 0 &0\\
0& 0 &0 & 0 &0\\
0& 0 &0 & 0 &0\\
\end{array}
\right],
\end{IEEEeqnarray*} 
\end{small}
meaning that the channel LLRs of VNs $3$, $5$ and $4$ are utilized in computation of state variables $7$, $8$ and $9$, respectively. 
\end{ex}
\begin{rem}\label{rowcolsystflood_rem}
While the general structure of the matrices involved in the state-space model of a LETS in row and column layered decoders are similar, the number of layers and the 
sub-matrices involved in different matrices are in general different for the two decoders.
\end{rem}

In the linear state-space model of a LETS, DE is used to obtain the distribution of external messages entering the LETS subgraph from the rest of the Tanner graph via mis-satisfied and unsatisfied CNs. In column layered decoding of QC-LDPC codes, the probability distribution of messages can be obtained based on the base graph of the code. In other words, the edges of the base graph are representatives of the cluster of edges between different types of VNs and CNs. In the base graph of a QC-LDPC code, within column layer $j$ of iteration $\ell$, Type-$i$ CN to Type-$j$ VN distribution denoted by $\psi_{\ell}^{[j\leftarrow i]}$ is obtained based on 
the distribution of messages passed to Type-$i$ CNs from other VNs with types $k \in N(i)\backslash j$. Depending on the type of these VNs as well as the updating order of the code layers, the VN types $k\in N(i)\backslash j$ are partitioned into two groups: 
1) Those that have already been updated in the previous layers of the current iteration $\ell$. This group pass the updated messages with distribution $\psi_{\ell}^{[k\rightarrow i]}$ to Type-$i$ CNs. 2) Those that belong to the subsequent layers of iteration $\ell$ and have not been updated yet. This group pass messages calculated 
at iteration $\ell-1$ with distribution $\psi_{\ell-1}^{[k\rightarrow i]}$. Following the procedure just described, the probability distribution of messages between different types of VNs and CNs at various iterations and layers are numerically calculated.\vspace{-.4cm}

\begin{figure}
\centering
\includegraphics[width=2.8in]{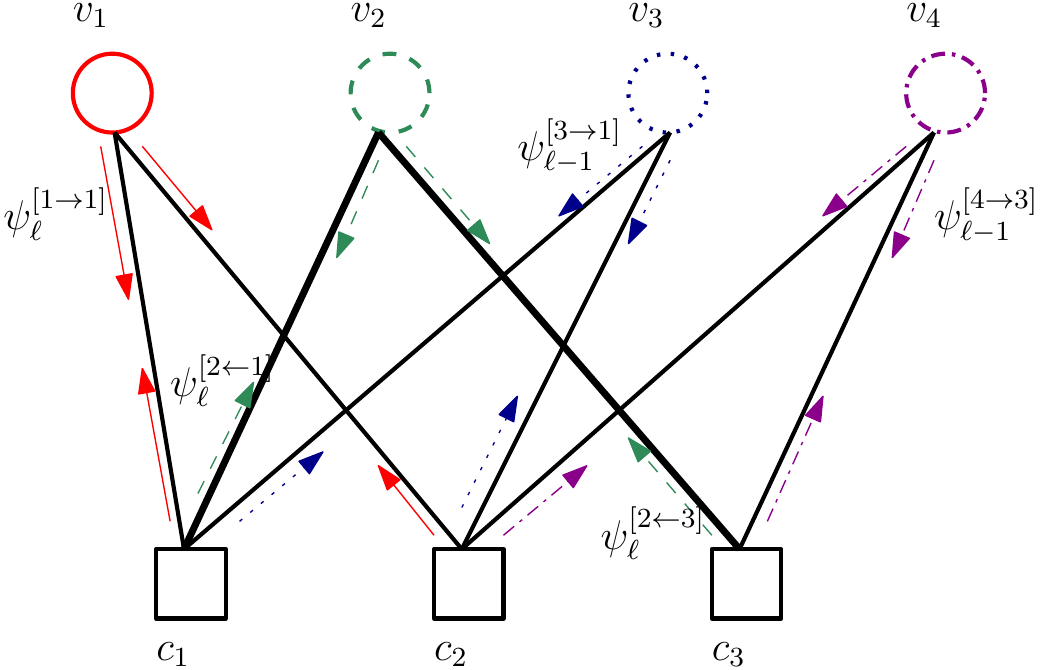}
\vspace{-0.5cm}
\caption{An example of a base graph with four column layers: Solid (red), dashed (green), dotted (blue) and dashed-dotted (purple) lines represent layers one to four, respectively.}
\label{BaseGraphColLay}
\end{figure}

\begin{ex}
Consider the base graph of Fig. \ref{BaseGraphColLay} with $n_b=4$ and $m_b=3$, and suppose a column layered schedule in which the four layers are updated 
starting from the leftmost VN of the base graph all the way to the rightmost VN. In Fig. \ref{BaseGraphColLay}, layers one to four are shown by 
solid (red), dashed (green), dotted (blue) and dashed-dotted (purple) lines, respectively.
The symbols next to the dashed arrows (on boldfaced edges) represent the probability distribution of CN to VN messages activated at 
the $2$nd layer of the $\ell$-th iteration. In the figure, we have also shown VN to CN distributions $\psi_{\ell}^{[1 \rightarrow 1]}$ and $\psi_{\ell-1}^{[3 \rightarrow 1]}$, 
updated in current and previous iterations, respectively, that are needed for the calculation of $\psi_{\ell}^{[1 \leftarrow 2]}$. 
Similarly, as the $4$-th layer has not been updated yet, $\psi_{\ell-1}^{[4 \rightarrow 3]}$ is used to calculate $\psi_{\ell}^{[2 \leftarrow 3]}$. 
\end{ex}\vspace{-.61cm}

In layered decoders, the failure rate of a LETS structure depends not only on the topology of the structure but also on the location of the TS within the Tanner graph of the code. The latter would affect the assignment of LETS's internal VNs to different layers (types), as well as the layer assignment of different external VNs connected to mis-satisfied and unsatisfied CNs of the LETS. In addition, a change in the location of a TS within the code's Tanner graph can change the type of the internal CNs of the TS. The collective information of layer (type) assignment of VNs and the types of CNs, just described, is referred to as \emph{TS layer profile} (TSLP) \cite{Ali-ITrow}. TSLP affects both the matrices in the state-space model of the LETS, and the distribution of the external messages entering the TS subgraph.

%

\subsection{Linear Gain Model for Mis-satisfied CNs}\label{linearGainColay}
\begin{figure}
\centering
\includegraphics[width=3.4in]{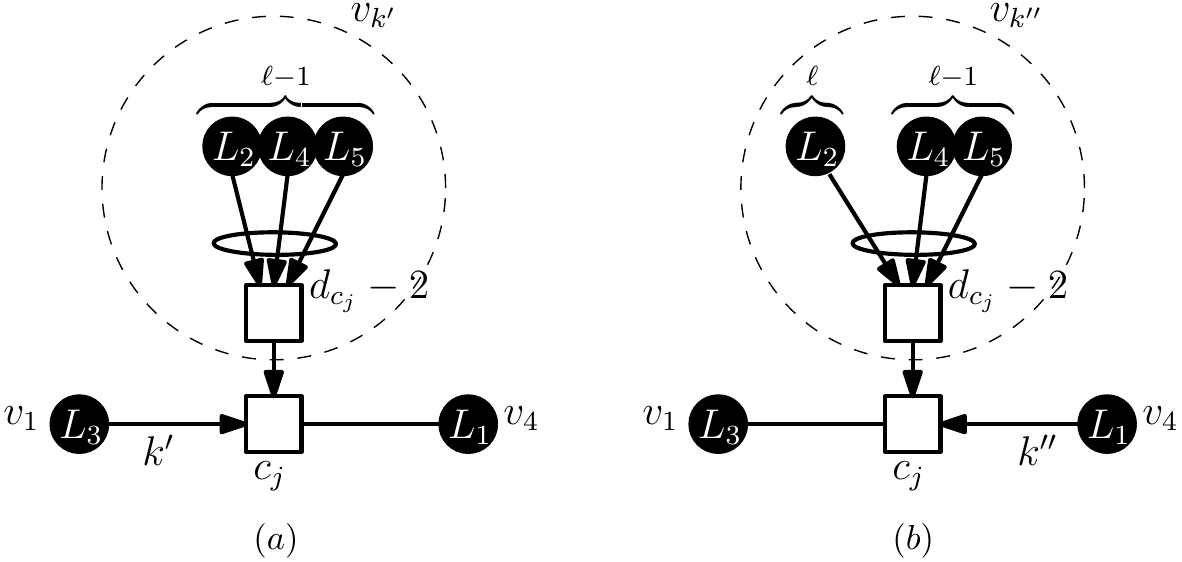}
\vspace{-0.5cm}
\caption{Two virtual VNs ($v_{k'}$ and $v_{k''}$) representing the effect of external VNs of the TS with respect to a mis-satisfied CN (from Fig. \ref{(5,3)collay_col}). The virtual VN depends on which output message of the CN inside the TS is updated: (a) $v_{k'}$ is used when message from $c_j$ to $v_4$ is updated (in $L_1$), (b) $v_{k''}$ is used when message from $c_j$ to $v_1$ is updated (in $L_3$). }
\label{mis_sat_53Col}
\end{figure}

In the state-space model of layered decoders, the impact of external connections of each mis-satisfied CN on the corresponding two state variables is modeled 
by a linear gain. These gains form the diagonal elements of the gain matrix, discussed in Section~\ref{SS_Col_subSec}. While for the row layered decoders, the gain for both state variables corresponding to each mis-satisfied CN is the same, for column layered decoders, the two gains can be different as discussed in the following.

For simplicity, suppose that the external connections of each mis-satisfied CN at each layer is represented by a virtual VN, as shown in Fig. \ref{mis_sat_53Col}. 
Consider the case in Fig. \ref{mis_sat_53Col}(a), where this virtual variable node is denoted by $v_{k'}$, and we are interested in calculating the linear gain 
affecting the message labeled by $k'$, as it is passed from $c_j$ to $v_4$ in the first layer of decoding. Then, the linear gain, $\bar{g}^{(\ell)}_{k'}$,
averaged over the probability density function $\hat{\psi}^{[k'\rightarrow j]}_\ell$ of the message from $v_{k'}$ to $c_j$ at iteration $\ell$, is given by
\begin{small}
\begin{equation}
\bar{g}^{(\ell)}_{k'}=\int_{-\infty}^{\infty}\tanh(\frac{\lambda}{2})\hat{\psi}^{[k'\rightarrow j]}_\ell(\lambda)d\lambda\:.
\label{g_layCol_I}
\end{equation}
\end{small}
Given the TSLP of a LETS, the function $\hat{\psi}^{[k'\rightarrow j]}_\ell$ can be calculated from the DE results for different mis-satisfied CNs at a specific layer. 
In Fig. \ref{mis_sat_53Col}, for example, 
$\hat{\psi}^{[k'\rightarrow j]}_\ell$ can be viewed as the CN to VN message distribution of a CN with degree $4$ (or $d_{c_j}-1$, in general) 
where the inputs are messages from Type-$2$, $4$ and $5$ VNs passed to Type-$1$ CNs ($c_j$ is a Type-1 CN).  
It is noted that unlike the case for a row layered decoder, the distribution of messages from the virtual VN to a mis-satisfied CN can be different,
depending on the column layer of the corresponding internal VNs, i.e., the virtual VN, and the distribution of messages from this node 
depend on the internal incoming message to a mis-satisfied CN that is being activated in a specific layer. This is explained in
Figs. \ref{mis_sat_53Col}(a) and (b), for the mis-satisfied CN $c_j$.
In Fig. \ref{mis_sat_53Col}(a), where the message $k'$ is activated at layer $L'_1=L_1$, none of the external VNs have been updated 
in the current iteration yet. However, in Fig. \ref{mis_sat_53Col}(b), where the message $k''$ is activated at layer $L'_2=L_3$, 
one of the external VNs that belongs to layer $L_2$ passes an updated message to $c_j$ while the others pass their information from the previous iteration, $\ell-1$. 
This implies that the distribution of messages from the virtual VN to the mis-satisfied CN is different for different state variables that are activated in different layers of the same iteration. Therefore, for each of the state variables (messages) passed to a mis-satisfied CN, a separate gain needs to be calculated.

Similar to~\cite{But_SS,Ali-ITrow}, Equation \eqref{g_layCol_I} is further modified by adding a polarity inversion effect into the model. 
Whenever an error occurs in the messages from the virtual VNs, $v_{k'}$, $k'=1,\dots,m_s$, the polarity of the corresponding state 
variables is altered. Denoting the probability of inversion by $P_{inv,\ell}^{[k'\rightarrow j]}$, we have
\begin{small}
\begin{equation}
P_{inv,\ell}^{[k'\rightarrow j]}=\int_{-\infty}^{0}\hat{\psi}^{[k'\rightarrow j]}_\ell(\lambda)d\lambda\:.
\label{Pinv_Colay}
\end{equation}
\end{small}\vspace{-.8cm}

To incorporate the polarity inversion into the model, similar to~\cite{But_SS,Ali-ITrow}, we modify the average linear gain to
\begin{small}
\begin{equation}
\bar{g}'^{(\ell)}_{k'}=(1-P_{inv,\ell}^{[k'\rightarrow j]})\bar{g}^{(\ell)}_{k'}\:.
\end{equation}  
\end{small}
As the iterations go on, the gains of different mis-satisfied CNs approach $1$.

In the following, we present an alternate approach to calculate the average gain of (\ref{g_layCol_I}). The new approach simplifies the calculations required for 
optimization of layers' order to minimize the error floor of column layered decoders.

Consider the case where a state variable $x_i$ is activated in a specific layer, and suppose that $x_i$ is the incoming message to a mis-satisfied CN $c_j$,
and that we wish to calculate the linear gain imposed on $x_i$ as $c_j$ passes it to an adjacent internal VN $v_{i'}$. For this scenario,
let the indices of external VNs to $c_j$ whose information have been updated at iterations $\ell-1$ and $\ell$ be denoted by the sets, $\overset{\triangleleft}{\tau}_i$ and $\overset{\triangleright}{\tau}_i$, respectively. Using equation \eqref{SPA_CN} to compute the message from $c_j$ to $v_{i'}$, we have\vspace{-.3cm}
\begin{small}
\begin{IEEEeqnarray*}{l}
L_{\ell}^{[i' \leftarrow j]}=2\tanh^{-1}\Bigg[\tanh{(\frac{x_{i}}{2})}
\prod_{k_1\in \overset{\triangleleft}{\tau}_i }\tanh\frac{{L_{\ell-1}^{[k_1 \rightarrow j]}}}{2}{\prod_{k_2 \in  \overset{\triangleright}{\tau}_i}\tanh\frac{{L_{\ell}^{[k_2 \rightarrow j]}}}{2}}\Bigg].\IEEEyesnumber
\label{ButlerGainApproach}
\end{IEEEeqnarray*} 
\end{small}
Considering the right hand side of the above equation as a function $f(x_i)$ of $x_i$, to obtain the linear gain, one needs to compute
the first derivative of $f(x_i)$ at $x_i=0$ ($f_{x}(0)$), and then take the expected value of such a derivative. We have
\begin{small}
\begin{IEEEeqnarray*}{l}
f_{x}(0)=\prod_{{k_1}\in \overset{\triangleleft}{\tau}_i }\tanh\frac{{L_{\ell-1}^{[k_1 \rightarrow j]}}}{2}{\prod_{{k_2} \in  \overset{\triangleright}{\tau}_i }\tanh\frac{{L_{\ell}^{[k_2 \rightarrow j]}}}{2}}\:.\IEEEyesnumber
\label{ButlerGainApproachFirstDeriv}
\end{IEEEeqnarray*}
\end{small}
The expected value of (\ref{ButlerGainApproachFirstDeriv}), assuming that the external messages entering the mis-satisfied CNs are independent, is given by 
\begin{small}
\begin{equation}
\bar{g}^{(\ell)}_{i}=\prod_{{k_1}\in \overset{\triangleleft}{\tau}_i }\overset{\triangleleft}\theta^{(\ell)}_{k_1}\prod_{{k_2} \in  \overset{\triangleright}{\tau}_i }\overset{\triangleright}\theta^{(\ell)}_{k_2},
\label{g_Colay}
\end{equation}
\end{small}\vspace{-0.2cm}
where the {\em partial VN to CN gains}, $\overset{\triangleleft}\theta^{(\ell)}_{k_1}$ and $\overset{\triangleright}\theta^{(\ell)}_{k_2}$, are defined as
\begin{small}
\begin{equation}
\overset{\triangleleft}\theta^{(\ell)}_{k_1}=\int_{-\infty}^{\infty}\tanh(\frac{\lambda}{2})\psi^{[k_1\rightarrow j]}_{\ell-1}(\lambda)d\lambda, \ \  {k_1}\in \overset{\triangleleft}{\tau}_i \:,
\label{g_partial_Colay_p}
\end{equation}
\vspace{-0.75cm}
\begin{equation}
\overset{\triangleright}\theta^{(\ell)}_{k_2}=\int_{-\infty}^{\infty}\tanh(\frac{\lambda}{2})\psi^{[k_2\rightarrow j]}_{\ell}(\lambda)d\lambda, \ \  {k_2} \in  \overset{\triangleright}{\tau}_i\:. 
\label{g_partial_Colay_c}
\end{equation}
\end{small}
Functions $\psi^{[k_1\rightarrow j]}_{\ell-1}$ and $\psi^{[k_2\rightarrow j]}_\ell$ are the VN to CN probability density functions corresponding to 
different external connections of a given mis-satisfied CN at iterations $\ell-1$ and $\ell$, respectively (they are different from 
the distribution of messages from virtual VNs, $\hat{\psi}^{[k'\rightarrow j]}_\ell$). 

Equation \eqref{g_Colay} will be used in section \ref{col_opt_Sec} to simplify the column layer scheduling optimization problem. 
To carry out the optimization, we use an approximate DE technique to reduce the complexity. However, changing the order of layers 
would affect the sets $\overset{\triangleleft}{\tau}_i$ and $\overset{\triangleright}{\tau}_i$ of each of the mis-satisfied CNs. As a result, 
if one were to use (\ref{g_layCol_I}) to calculate the gains, a new distribution would need to be calculated each time. To tackle this problem, 
we first calculate the partial gains \eqref{g_partial_Colay_p} and \eqref{g_partial_Colay_c} in advance, and then
use them in  \eqref{g_Colay} in accordance with $\overset{\triangleleft}{\tau}_i$ and $\overset{\triangleright}{\tau}_i$, to calculate the gains of different mis-satisfied CNs. 
We note that, for a specific VN to CN message, we have\vspace{-.3cm}
\begin{small}
 \begin{equation}
 \overset{\triangleleft}\theta^{(\ell)}_{k}=\overset{\triangleright}\theta^{(\ell-1)}_{k}.
\end{equation}   
\end{small}
\begin{ex}
Suppose that the external VNs of CN $c_j$ in Fig. \ref{mis_sat_53Col} are labeled with the same indices of their corresponding layers as $v'_2$, $v'_4$, and $v'_5$. 
Then the VN to CN distributions $\psi_{\ell-1}^{[2 \rightarrow j]}$, $\psi_{\ell-1}^{[4 \rightarrow j]}$ and $\psi_{\ell-1}^{[5 \rightarrow j]}$ are needed to 
calculate the virtual VN message distribution $\hat{\psi}^{[k'\rightarrow j]}_\ell$. Also, the VN to CN distributions $\psi_{\ell}^{[2 \rightarrow j]}$, $\psi_{\ell-1}^{[4 \rightarrow j]}$ and $\psi_{\ell-1}^{[5 \rightarrow j]}$ are used to calculate the virtual VN message distribution $\hat{\psi}^{[k''\rightarrow j]}_\ell$. 
Correspondingly, the partial VN to CN gains for $x_{k'}$ and $x_{k''}$ are obtained as
\begin{small}
  \begin{equation}\label{ex1partialgain}
\bar{g}^{(\ell)}_{k'}=\overset{\triangleleft}\theta^{(\ell)}_{2}\overset{\triangleleft}\theta^{(\ell)}_{4}\overset{\triangleleft}\theta^{(\ell)}_{5},
\end{equation}%
\end{small}
\vspace{-0.8cm}
\begin{small}
\begin{equation}\label{ex2partialgain}
\bar{g}^{(\ell)}_{k''}=\overset{\triangleright}\theta^{(\ell)}_{2}\overset{\triangleleft}\theta^{(\ell)}_{4}\overset{\triangleleft}\theta^{(\ell)}_{5},
\end{equation}
\end{small}
where $\overset{\triangleleft}{\tau}_{k'}=\{2,4,5\}$, $\overset{\triangleright}{\tau}_{k'} = \emptyset$,  $\overset{\triangleleft}{\tau}_{k''}=\{4,5\}$ and $\overset{\triangleright}{\tau}_{k''}=\{2\}$. 
\end{ex}
\subsection{Failure Probability}
The probability of a TS failure in column layered decoders can be obtained by the same general approach as flooding and row layered decoders. 
First, the non-recursive form of the state vector as a function of model inputs is derived. The resulted equation, then, is used to 
obtain an error indicator function. Finally, using Gaussian approximation, the failure probability of the TS is estimated. The details are given below. 

To derive the non-recursive equation of the state variables, we first need to define some matrices.
The ordered multiplication of the scaled version of transition matrices of different layers, $\mathfrak{G}_j^{(\ell)}\mathcal{A}_j$, from the maximum layer $J$ down to the $k$-th layer, $k \geq 1$, is defined as
\begin{small}
\begin{equation}\label{A_jdowntok_l_ColLay}
\tilde{\mathbf{A}}_{J \rightarrow k}^{(\ell)}=(\mathfrak{G}_J^{(\ell)}\mathcal{A}_J)  (\mathfrak{G}_{J-1}^{(\ell)}\mathcal{A}_{J-1}) \dots (\mathfrak{G}_k^{(\ell)}\mathcal{A}_k).
\end{equation}
\end{small}
The superscript $(\ell)$ shows the iteration dependency of this matrix due to time-variant gain matrices. 
By using $\tilde{\mathbf{A}}_{J \rightarrow k}^{(\ell)}$, three other matrices are defined as  
\begin{small}
\begin{equation}
\tilde{\mathbf{B}}^{(\ell)}=\tilde{\mathbf{A}}_{J \rightarrow 2}^{(\ell)}\mathfrak{G}_1^{(\ell)}\mathfrak{B}_1+
\tilde{\mathbf{A}}_{J \rightarrow 3}^{(\ell)}\mathfrak{G}_2^{(\ell)}\mathfrak{B}_2+\dots+\mathfrak{G}_J^{(\ell)}\mathfrak{B}_J,
\end{equation}
\begin{equation}
\overset{\triangleright}{\mathbf{ B}}_{ex}^{(\ell)}=\tilde{\mathbf{A}}_{J \rightarrow 2}^{(\ell)}\mathfrak{G}_1^{(\ell)}\overset{\triangleright}{\mathfrak{B}}_{ex1}+
\tilde{\mathbf{A}}_{J \rightarrow 3}^{(\ell)}\mathfrak{G}_2^{(\ell)}\overset{\triangleright}{\mathfrak{B}}_{ex2}+\dots+\mathfrak{G}_J^{(\ell)}\overset{\triangleright}{\mathfrak{B}}_{exJ},
\end{equation}
\begin{equation}\label{Bex_backtriangel_lColay}
\overset{\triangleleft}{\mathbf{ B}}_{ex}^{(\ell)}=\tilde{\mathbf{A}}_{J \rightarrow 2}^{(\ell)}\mathfrak{G}_1^{(\ell)}\overset{\triangleleft}{\mathfrak{B}}_{ex1}+
\tilde{\mathbf{A}}_{J \rightarrow 3}^{(\ell)}\mathfrak{G}_2^{(\ell)}\overset{\triangleleft}{\mathfrak{B}}_{ex2}+\dots+\mathfrak{G}_J^{(\ell)}\overset{\triangleleft}{\mathfrak{B}}_{exJ}.
\end{equation}
\end{small}
Using the above matrices, the non-recursive form of the state vector at the end of iteration $\ell$ is derived as   \vspace{-.3cm}
\begin{small}
\begin{IEEEeqnarray*}{l}
\label{ss_lay_mainColay}
\tilde{\mathbf{x}}^{(\ell,J)}=\sum_{i=1}^\ell \big(\prod_{j=i+1}^{\underrightarrow{\ell}}\tilde{\mathbf{A}}_{J \rightarrow 1}^{(j)}\big)\tilde{\mathbf{B}}^{(i)}\L
\ +\sum_{i'=1}^\ell \big(\prod_{j'=i'+1}^{\underrightarrow{\ell}}\tilde{\mathbf{A}}_{J \rightarrow 1}^{(j')}\big)\big(\overset{\triangleleft}{\mathbf{ B}}_{ex}^{(i')}\L_{ex}^{(i'-1)}+\overset{\triangleright}{\mathbf{ B}}_{ex}^{(i')}\L_{ex}^{(i')}\big)\:,
\IEEEyesnumber
\end{IEEEeqnarray*}
\end{small}
which is a function of the inputs from unsatisfied CNs as well as the channel.
The right arrow on top of the product sign indicates that the matrix multiplication is applied from the left. For example,\vspace{-.3cm}
\begin{small}
\begin{equation*}
\prod_{j=i+1}^{\underrightarrow{\ell}}\tilde{\mathbf{A}}_{J \rightarrow 1}^{(j)}=\tilde{\mathbf{A}}_{J \rightarrow 1}^{(\ell)}\tilde{\mathbf{A}}_{J \rightarrow 1}^{(\ell-1)}\dots \tilde{\mathbf{A}}_{J \rightarrow 1}^{(i+1)}.
\end{equation*} 
\end{small}
As mentioned before, the mis-satisfied CN average gains tend to $1$ as the iterations tend to infinity. So, we have
\begin{small}
\begin{equation}\label{Atild_iterDepend_AtildIterIndepend}
\displaystyle{\lim_{\ell \to \infty}}\tilde{\mathbf{A}}_{J \rightarrow 1}^{(\ell)}=\mathcal{A}_J\mathcal{A}_{J-1} \dots  \mathcal{A}_1=\tilde{\mathbf{A}}_{J \rightarrow 1}.\\
\end{equation} 
\end{small}
The matrix $\tilde{\mathbf{A}}_{J \rightarrow 1}$, which is not iteration dependent, is referred to as the {\em transition matrix of the layered decoder}, 
and plays a crucial role in determining the growth rate of erroneous messages within the LETS subgraph.

We note that, the systematic forms of transition matrices for row and column layered decoders are similar (both have a similar block structure).
As a result, all the spectral properties of a row layered transition matrix, proved in [my paper], are also valid for the transition matrix of a column layered decoder. Some of the notable spectral properties of $\tilde{\mathbf{A}}_{J \rightarrow 1}$ are:
\begin{itemize}
\item[(a)] Matrix $\tilde{\mathbf{A}}_{J \rightarrow 1}$ of a LETS is reducible.
\item[(b)] $\tilde{\mathbf{A}}_{J \rightarrow 1}$ has a positive dominant eigenvalue, $\tilde{r}$. The value of $\tilde{r}$ is equal to the spectral radius 
of $\tilde{\mathbf{A}}_{J \rightarrow 1}$, and the left and right eigenvectors, $\tilde{\mathbf{w}}_1$ and $\tilde{\mathbf{u}}_1$, corresponding to 
$\tilde{r}$ are both non-negative and non-zero. 
\item[(c)] For any LETS structure, the dominant eigenvalue $\tilde{r}$ of a layered transition matrix, is always greater than or equal 
to the corresponding dominant eigenvalue $r$ of the transition matrix for the flooding schedule, i.e., $\tilde{r}\geq r$. 
\item[(d)] For any LETS structure, the dominant eigenvalue of the column layered transition matrix is invariant to any cyclic shift of the layers' order. 
\item[(e)] For any LETS structure, the dominant eigenvalue of the column layered transition matrix is invariant to the 
reversing of the layers' order.\footnote{The proof of this property, which is slightly different from that of the row layered decoder~\cite{Ali-ITrow}, is provided in Appendix. } 
\item[(f)] For a LETS structure with $J$ column layers $(J\geq 3)$, the number of distinct dominant eigenvalues of the column layered transition matrix 
corresponding to different permutations of the layers is upper bounded by $\frac{(J-1)!}{2}$.
\end{itemize} 

Similar to the approach of~\cite{Ali-ITrow, But_SS}, here, we also use the projection of the state vector of \eqref{ss_lay_mainColay} onto the dominant left eigenvector of $\tilde{\mathbf{A}}_{J \rightarrow 1}$ as the error indicator function $\tilde{\beta}^{(\ell)}$:
\begin{small}
\begin{IEEEeqnarray*}{l}
\tilde{\beta}^{(\ell)}\triangleq \mathbf{\tilde{w}}_1^T\tilde{\mathbf{x}}^{(\ell,J)}=\boldsymbol{\gamma}_{ch}^T\L
\ +\sum_{i'=1}^\ell \big({\overset{\triangleleft}{\boldsymbol{\gamma}}_{ex}^{(i')T}}\L_{ex}^{(i'-1)}+\overset{\triangleright}{\mathbf{ \boldsymbol{\gamma}}}_{ex}^{(i')T}\L_{ex}^{(i')}\big)\:.
\label{eq34}
\IEEEyesnumber
\end{IEEEeqnarray*}
\end{small}\vspace{-0.1cm}
The vectors $\boldsymbol{\gamma}_{ch}^T$, $\overset{\triangleleft}{\boldsymbol{\gamma}}_{ex}^{(i')T}$ and $\overset{\triangleright}{\boldsymbol{\gamma}}_{ex}^{(i')T}$ in (\ref{eq34}) have lengths $a$, $b$ and $b$, respectively, and are defined as  
\begin{small}
\begin{equation}\label{gamma_ch_TColay}
\boldsymbol{\gamma}_{ch}^T=\mathbf{\tilde{w}}_1^T\sum_{i=1}^\ell \big(\prod_{j=i+1}^{\underrightarrow{\ell}}\tilde{\mathbf{A}}_{J \rightarrow 1}^{(j)}\big)\tilde{\mathbf{B}}^{(i)}\:,
\end{equation}\vspace{-.3cm}
\begin{equation}\label{gamaExleftColay}
\overset{\triangleleft}{\boldsymbol{\gamma}}_{ex}^{(i')T}=\mathbf{\tilde{w}}_1^T\big(\prod_{j'=i'+1}^{\underrightarrow{\ell}}\tilde{\mathbf{A}}_{J \rightarrow 1}^{(j')}\big)\overset{\triangleleft}{\mathbf{ B}}_{ex}^{(i')} \ \ \ \ \  i'=1,\dots,\ell\:,
\end{equation}\vspace{-.3cm}
\begin{equation}\label{gamaExRightColay}
\overset{\triangleright}{\boldsymbol{\gamma}}_{ex}^{(i')T}=\mathbf{\tilde{w}}_1^T\big(\prod_{j'=i'+1}^{\underrightarrow{\ell}}\tilde{\mathbf{A}}_{J \rightarrow 1}^{(j')}\big)\overset{\triangleright}{\mathbf{ B}}_{ex}^{(i')}\ \ \ \ \  i'=1,\dots,\ell\:.
\end{equation}
\end{small}
The mean and the variance of $\tilde{\beta}^{(\ell)}$ are calculated as 
\begin{small}
\begin{IEEEeqnarray*}{lCl"s}\label{Mean_beta_layeredColay}
\mathbb{E}[\tilde{\beta}^{(\ell)}]=\frac{2}{\sigma^2_{ch}}\sum_{k=1}^a(\boldsymbol{\gamma}_{ch}^T)_k 
+\sum_{i'=1}^{\ell-1} \big({\overset{\triangleleft}{\boldsymbol{\gamma}}_{ex}^{(i'+1)T}}+\overset{\triangleright}{\mathbf{ \boldsymbol{\gamma}}}_{ex}^{(i')T}\big)\mathbf{m}_{ex}^{(i')}+ \overset{\triangleright}{\mathbf{ \boldsymbol{\gamma}}}_{ex}^{(\ell)T}\mathbf{m}_{ex}^{(\ell)}\:,
\IEEEyesnumber
\end{IEEEeqnarray*}\vspace{-.5cm}
\begin{IEEEeqnarray*}{lCl"s}
\label{Var_beta_layeredColay}
\mathbb{VAR}[\tilde{\beta}^{(\ell)}]=\frac{4}{\sigma^2_{ch}}\sum_{k=1}^a(\boldsymbol{\gamma}_{ch}^T)_k^2 
+\sum_{i'=1}^{\ell-1} \big({\overset{\triangleleft}{\boldsymbol{\gamma}}_{ex}^{(i'+1)T}}+\overset{\triangleright}{\mathbf{ \boldsymbol{\gamma}}}_{ex}^{(i')T}\big)\mathbf{\Sigma}_{ex}^{(i')}\big({\overset{\triangleleft}{\boldsymbol{\gamma}}_{ex}^{(i'+1)}}+\overset{\triangleright}{\mathbf{ \boldsymbol{\gamma}}}_{ex}^{(i')}\big)+
\overset{\triangleright}{\mathbf{ \boldsymbol{\gamma}}}_{ex}^{(\ell)T}\mathbf{\Sigma}_{ex}^{(\ell)}\overset{\triangleright}{\mathbf{ \boldsymbol{\gamma}}}_{ex}^{(\ell)}\:.
\IEEEyesnumber
\end{IEEEeqnarray*}
\end{small}
In the above equations, the symbol $(.)_k$ is used to represent the $k$-th element of the vector inside the parentheses. Parameters $m_{ch}$ and $\sigma^2_{ch}$ are scalars representing the mean and variance of the channel LLRs. The $b\times1$ vector $\mathbf{m}_{ex}^{(i')}$ includes the 
mean of the external inputs from unsatisfied CNs at iteration $i'$. The $b\times b$ matrix $\mathbf{\Sigma}_{ex}^{(i')}$ is the covariance matrix 
of the inputs from unsatisfied CNs at iteration $i'$. Since it is assumed that different external inputs are independent, $\mathbf{\Sigma}_{ex}^{(i')}$
 is a diagonal matrix. The mean and variance of the inputs from unsatisfied CNs are calculated based on DE. Also, the mis-satisfied CN gains 
 are calculated based on DE, using the method of Section \ref{linearGainColay}. These gains are utilized in iteration-dependent gain matrices involved in equations \eqref{A_jdowntok_l_ColLay} to \eqref{Bex_backtriangel_lColay}. The vector $\boldsymbol{\gamma}_{ch}^T$ in \eqref{gamma_ch_TColay} as well as the set of vectors $\overset{\triangleleft}{\boldsymbol{\gamma}}_{ex}^{(i')T}$, $\overset{\triangleright}{\boldsymbol{\gamma}}_{ex}^{(i')T}$, $i'=1,\dots,\ell$, in \eqref{gamaExleftColay} and \eqref{gamaExRightColay} are then calculated and plugged into equations \eqref{Mean_beta_layeredColay} and \eqref{Var_beta_layeredColay} to find the mean and variance of the error indicator function. Finally, assuming that $\tilde{\beta}^{(\ell)}$  is Gaussian, the failure probability of a LETS, $\mathcal{S}$, is obtained by
 \begin{small}
\begin{equation}\label{P_fail_LayeredColay}
{P_e(\mathcal{S})=\lim_{\ell\to\infty}Pr\{\tilde{\beta}^{(\ell)}<0\}=\lim_{l\to\infty}Q\bigg(\frac{\mathbb{E}[\tilde{\beta}^{(\ell)}]}{\sqrt{\mathbb{VAR}[\tilde{\beta}^{(\ell)}]}}\bigg)\:,}
\end{equation}
\end{small}
where $Q(x)=\int_x^{+\infty} \frac{1}{\sqrt{2\pi}} e^{-t^2/2} dt$. In practice, the above limit converges fast within only a few iterations.

In order to estimate the error floor of an LDPC code under column layered SPA, we first start by identifying the LETS classes that may have non-negligible contribution to the error floor. We then partition the LETSs within each class into subsets of isomorphic TSs, i.e., all the TSs within one subset are isomorphic, but any two TSs that do not belong to the same subset are non-isomorphic. Finally, we partition the TSs within each subset into {\em groups} in accordance with their TSLP, i.e., all TSs within a 
{\em TSLP group} are isomorphic and have the same TSLP. 
Using the union bound, we then estimate the error floor as
\begin{small}
 \begin{equation}
P_f \approx \sum_i \Upsilon_i P_e(\mathcal{S}_i),
\end{equation}
\end{small}
where $\mathcal{S}_i$ is the representative of the $i$-th TSLP group whose size is $\Upsilon_i$. \vspace{-0.2cm}
\begin{rem}
For each subset of isomorphic structures within a specific LETS class in an LDPC code, the number and size of distinct TSLP groups are the same for column and row layered schedules.
\end{rem}\vspace{-1cm}

\section{Optimization of Column Layered Schedule}\label{col_opt_Sec}
 
Generally, changing the column layered schedule by modifying the order in which different column layers are updated in each iteration of the decoding process
would change the TSLP of a LETS. As a result, the state-space model matrices, in general, and the transition matrix $\tilde{\mathbf{A}}_{J \rightarrow 1}$, in particular, 
would change. In addition, by the change in TSLP, the distribution of external messages entering the TS would change. Modifying the column layered schedule 
therefore can change the failure rate of a LETS as well as the error floor of the code. In this section, we use the state-space model developed in the 
previous section to study the effect of column layered schedule on the error floor and to find a schedule that minimizes the error floor.



One of the most computationally expensive tasks in using the state-space model to estimate the error floor is the implementation of DE, based on which the inputs of the model are determined. 
Consider a QC-LDPC code with an $m_b \times n_b$ base matrix. 
For such a base matrix, there exist $n_b!=n_b\times (n_b-1)\times \dots \times 1$ different column permutations, each corresponding to a different column layered schedule.
In the search for an optimal schedule, it would thus be too complex to find the error floor of each candidate schedule through an accurate analysis of the error floor using the state-space model and based on an accurate implementation of DE. 
In an accurate DE implementation for the column layered decoders, the probability distributions of CN to VN and VN to CN messages 
passing on different edges of the base graph are computed. In such a computation, the order in which different messages (distributions) are updated must be taken into account. Such an order, in turn, depends on the layered schedule.  
To simplify the search among the ${n_b}!$ different column layered schedules, rather than the accurate derivation of the above distributions for each schedule, 
we select one arbitrary schedule, say the one corresponding to the original order of the columns, and then at each iteration $\ell$ and for each layer $j$, 
we only derive the distributions of CN to VN messages and VN to CN messages for that schedule. Subsequently, we take the average 
of such distributions, denoted by $\stackrel{\leftarrow}{\bar{\psi}_{\ell}^j}$ and $\stackrel{\rightarrow}{\bar{\psi}_{\ell}^j}$, respectively, 
where the average is taken over all the corresponding distributions within column layer $L_j$. 
These average distributions are then used to represent all the CN to VN and VN to CN distributions in the $j$th layer of decoding regardless of the schedule. \vspace{-.4cm}
\begin{ex}
As an example, for the base graph of Fig. \ref{BaseGraphColLay}, the distribution $\stackrel{\leftarrow}{\bar{\psi}_{\ell}^2}$ is the average of $\psi_{\ell}^{[2\leftarrow 1]}$ and $\psi_{\ell}^{[2\leftarrow 3]}$, and $\stackrel{\rightarrow}{\bar{\psi}_{\ell}^2}$ is the average of $\psi_{\ell}^{[2\rightarrow 1]}$ and $\psi_{\ell}^{[2\rightarrow 3]}$ (the last two distributions are not shown in the figure). 
\end{ex}\vspace{-.4cm}
In the next step, by using the average distribution of VN to CN messages for each layer $j$, $\stackrel{\rightarrow}{\bar{\psi}_{\ell}^j}$, 
a partial VN to CN gain for that layer  can be obtained using an equation similar to~\eqref{g_partial_Colay_p} and \eqref{g_partial_Colay_c}.
Then, based on the TSLP of a LETS, for a specific permutation of column layers, equation \eqref{g_Colay} can be utilized 
to obtain the multiplicative gains of various mis-satisfied CNs. 
To further reduce the complexity, the polarity inversion coefficient is not considered in the gain calculations for the optimization process. 
We note that the advantage of using \eqref{g_Colay} is that the partial gains related to different layers can be 
computed prior to the optimization process and different schedules would only affect the parameters $\overset{\triangleleft}{\tau}_i$ and $\overset{\triangleright}{\tau}_i$ related to different state-variables passing through various mis-satisfied CNs. 
With respect to unsatisfied CNs, the average CN to VN distributions $\stackrel{\leftarrow}{\bar{\psi}_{\ell}^j}$ or $\stackrel{\leftarrow}{\bar{\psi}_{\ell-1}^j}$ 
are used to compute the unsatisfied CN input parameters. 
\begin{ex}
Consider the $(155,64)$ Tanner code and its $(5,3)$ LETS shown in Fig.~\ref{(5,3)collay_col}. This code has five column layers, with the corresponding 
average VN to CN distributions $\stackrel{\rightarrow}{\bar{\psi}_{\ell}^1}$, $\stackrel{\rightarrow}{\bar{\psi}_{\ell}^2}$, $\stackrel{\rightarrow}{\bar{\psi}_{\ell}^3}$, $\stackrel{\rightarrow}{\bar{\psi}_{\ell}^4}$ and $\stackrel{\rightarrow}{\bar{\psi}_{\ell}^5}$. Based on these distributions, for each column layer, a VN to CN partial gain is computed. In relation to the mis-satisfied CN of the LETS shown in Fig. \ref{mis_sat_53Col}, the obtained partial gains are plugged into equations \eqref{ex1partialgain} and \eqref{ex2partialgain} to derive the corresponding gains at layers $L'_1=L_1$ and $L'_2=L_3$, respectively.

Now, consider the unsatisfied CN $c_2$ in Fig. \ref{(5,3)collay_col}. The CN to VN message from $c_2$ to $v_4$, is used in the model when the state 
variables labeled $9$ and $12$ are updated. These state variables are activated in the column layers $L'_2=L_3$ and $L'_3=L_5$, respectively. 
As both of these layers are updated after $L'_1=L_1$, which is the column layer of $v_4$, 
the average distribution $\stackrel{\leftarrow}{\bar{\psi}_{\ell}^1}$ (not $\stackrel{\leftarrow}{\bar{\psi}_{\ell-1}^1}$) is used 
within the model to obtain the unsatisfied input parameters (mean and variance) corresponding to $c_2$. 
\end{ex}

In QC-LDPC codes, we often have $n_b > m_b$, and $n_b! >> m_b!$. (For example, the WiMAX code used in this paper, denoted by $\mathcal{C}_2$ in the next section, 
has a $6\times 24$ base matrix.) This implies that for column layered decoders, unlike the row layered decoders, an exhaustive search 
among different column layer permutations may not be feasible.
In [my paper], it was shown that the failure probability of a LETS structure, on average, decreases with the reduction of the dominant eigenvalue of the transition matrices corresponding to different schedules. We have observed a similar trend in column layered decoders. To reduce the size of the search space, we thus focus on schedules 
that result in the smallest dominant eigenvalue for transition matrices of the most dominant LETS structure of the code. We then perform a search among such candidate schedules by using the approximate DE, described above, to find a few that have lower error floors. We then apply the exact DE within the linear state-space model
to these few candidate schedules to find the one that has the lowest error floor.



\section{Simulation Results} \label{SimulaColSection}
\begin{figure*}

  \centering  
 \subfloat[Exponent matrix of $\mathfrak{C}_1$: $(640, 192)$ QC-LDPC code with lifting degree $64$.]{\label{C1_640par}
  $ 
  \scriptsize{
\begin{array}{|c| c |c |c |c |c |c| c| c| c|}
\hline
8&-1&26&62&-1&-1&59&19&-1&60\\
\hline
51&9&50&-1&39&-1&4&-1&25&26\\
\hline
-1&32&10&7&56&52&41&55&61&41\\\hline
46&11&-1&43&-1&63&8&51&37&-1\\\hline
47&7&7&50&49&53&-1&12&-1&-1\\\hline
31&-1&-1&-1&31&38&-1&-1&23&48\\\hline
-1&25&21&56&59&30&27&23&27&18\\\hline
\end{array}}
$}\\
\subfloat[Exponent matrix of $\mathfrak{C}_2$: $(576, 432)$ WiMAX QC-LDPC code with lifting degree 24.]{\label{C2_576par}
$\scriptsize{
\begin{array}{|c |c |c |c |c |c |c| c| c |c |c| c| c| c |c| c| c| c| c| c| c| c| c| c|}
\hline
-1&20&-1&7&-1&-1&3&6&4&-1&-1&21&7&13&19&23&5&23&0&0&-1&-1&-1&-1\\\hline
10&-1&3&17&8&-1&-1&-1&-1&17&10&2&9&10&8&14&9&6&-1&0&0&-1&-1&-1\\\hline
-1&-1&5&-1&-1&15&9&-1&17&16&-1&9&1&18&11&7&15&1&20&-1&0&0&-1&-1\\\hline
16&0&-1&-1&15&-1&-1&0&12&-1&20&3&23&2&21&9&3&4&-1&-1&-1&0&0&-1\\\hline
-1&13&15&20&-1&6&18&-1&-1&-1&-1&21&19&0&0&18&15&6&-1&-1&-1&-1&0&0\\\hline
19&-1&-1&-1&3&7&-1&8&-1&18&7&17&21&21&6&16&2&22&0&-1&-1&-1&-1&0\\\hline
\end{array}
}
$
}
\caption{QC-LDPC codes used for simulations. The entries of the matrices, that are not equal to $-1$, represent the right circular shift of the identity matrix to create the corresponding block of the parity-check matrix. The $-1$ entries represent zero blocks.}
\label{par_used_lay}
\end{figure*}
In this section, the performance of our proposed model is investigated via simulations. We consider two QC-LDPC codes, $\mathcal{C}_1$ and $\mathcal{C}_2$,
where $\mathcal{C}_1$ is a rate-$0.3$ variable-regular $(640,192)$ code with $d_v=5$ and irregular CN degrees \cite{Ryan1} and 
$\mathcal{C}_2$ is a rate-$0.75$ irregular $(576,432)$ code used in WiMAX standard \cite{wimax_standard}. The exponent matrices of both codes are provided in Fig.~\ref{par_used_lay}.

We first investigate the effect of column block permutations on the error floor of $\mathcal{C}_1$. Code $\mathcal{C}_1$ has $10$ column layers, each layer corresponding to one of the column blocks of the parity-check matrix in Fig. \ref{C1_640par}. We label these column blocks with numbers $1$ to $10$ based on their original order in Fig. \ref{C1_640par}, i.e., the first column block is labeled by $1$, the second by $2$, and so on. 
Different schedules can then be represented with different permutations of numbers from $1$ to $10$.
For example, permutation $(1, 2, 3, 4, 5, 6, 7, 8, 9, 10)$ refers to the updating schedule of column layers from left to right  
(in the same order as they appear in Fig. \ref{C1_640par}).

\begin{figure*}
\centering
\includegraphics[width=6.6in]{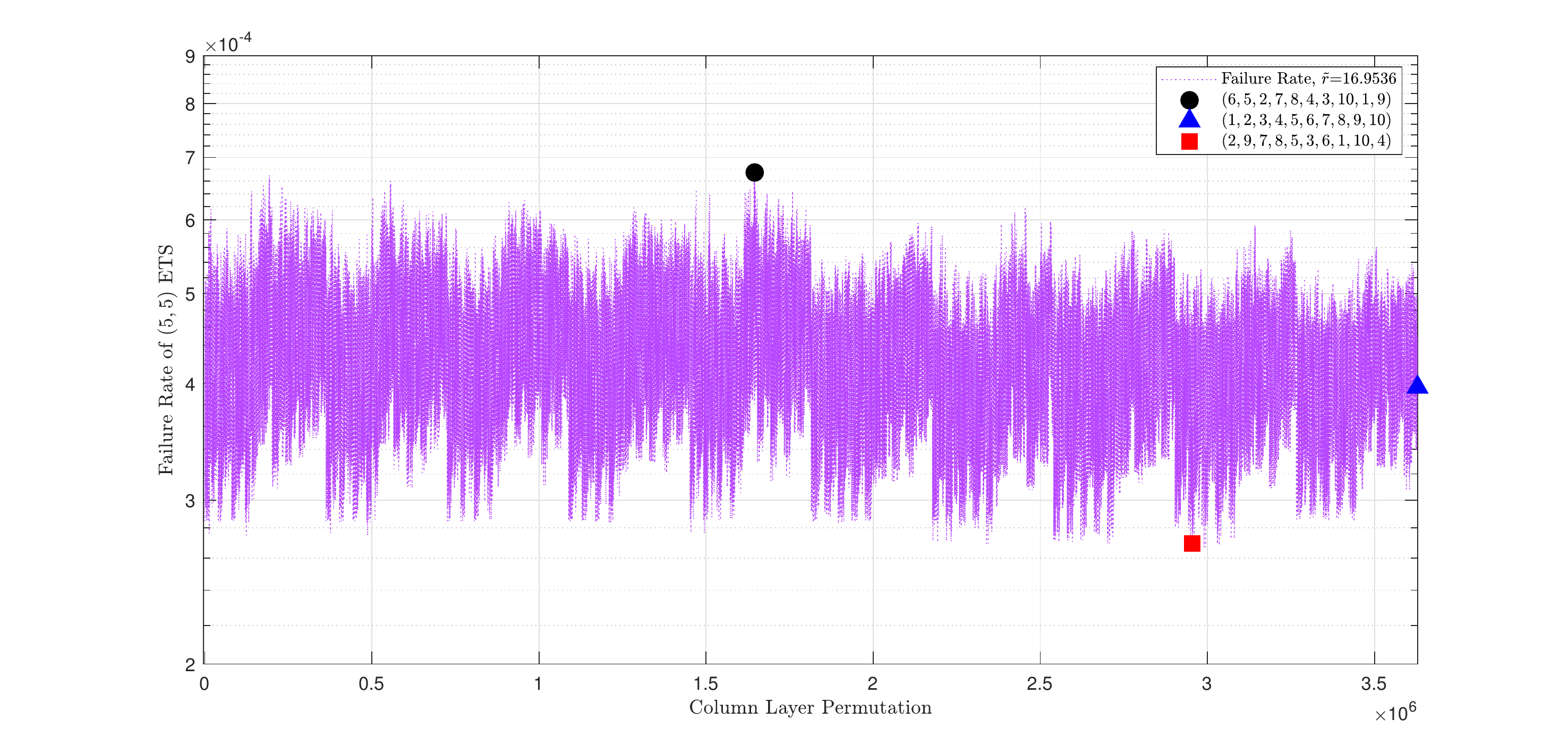}
\vspace{-1.4cm}
\caption{The approximate failure rate of $(5, 5)$ LETS of $\mathcal{C}_1$ for various column layered schedules at $E_b/N_0=6$ dB. All the schedules result in the same dominant eigenvalue of the layered transition matrix, $\tilde{r}=16.9536$. The saturation level is $15.75$.  }
\label{colOptimization55}
\end{figure*}

The dominant LETS of $\mathcal{C}_1$ is a $(5,5)$ structure with multiplicity $64$. All the $64$ instances of the $(5,5)$ LETS have the same TSLP. 
The total number of possible column layer permutations for $\mathcal{C}_1$ is $10!=3628800$. 
These permutations result in different transition matrices for the $(5,5)$ LETS within the state-space model. 
However, unlike the row layered schedules of this code presented in [my paper] where different row schedules produced $8$ different dominant eigenvalues, 
all the column layered schedules of this code result in transition matrices whose dominant eigenvalues are equal to $\tilde{r}=16.9536$. 
By using a single application of approximate DE, as discussed in Section \ref{col_opt_Sec}, we approximate the failure rate of the $(5,5)$ LETS for different 
schedules. These results at SNR of $6$ dB and saturation level of $15.75$ are provided in Fig. \ref{colOptimization55} for all the schedules.

\begin{figure}
\vspace{-0.6cm}
\centering
\includegraphics[width=3.5in]{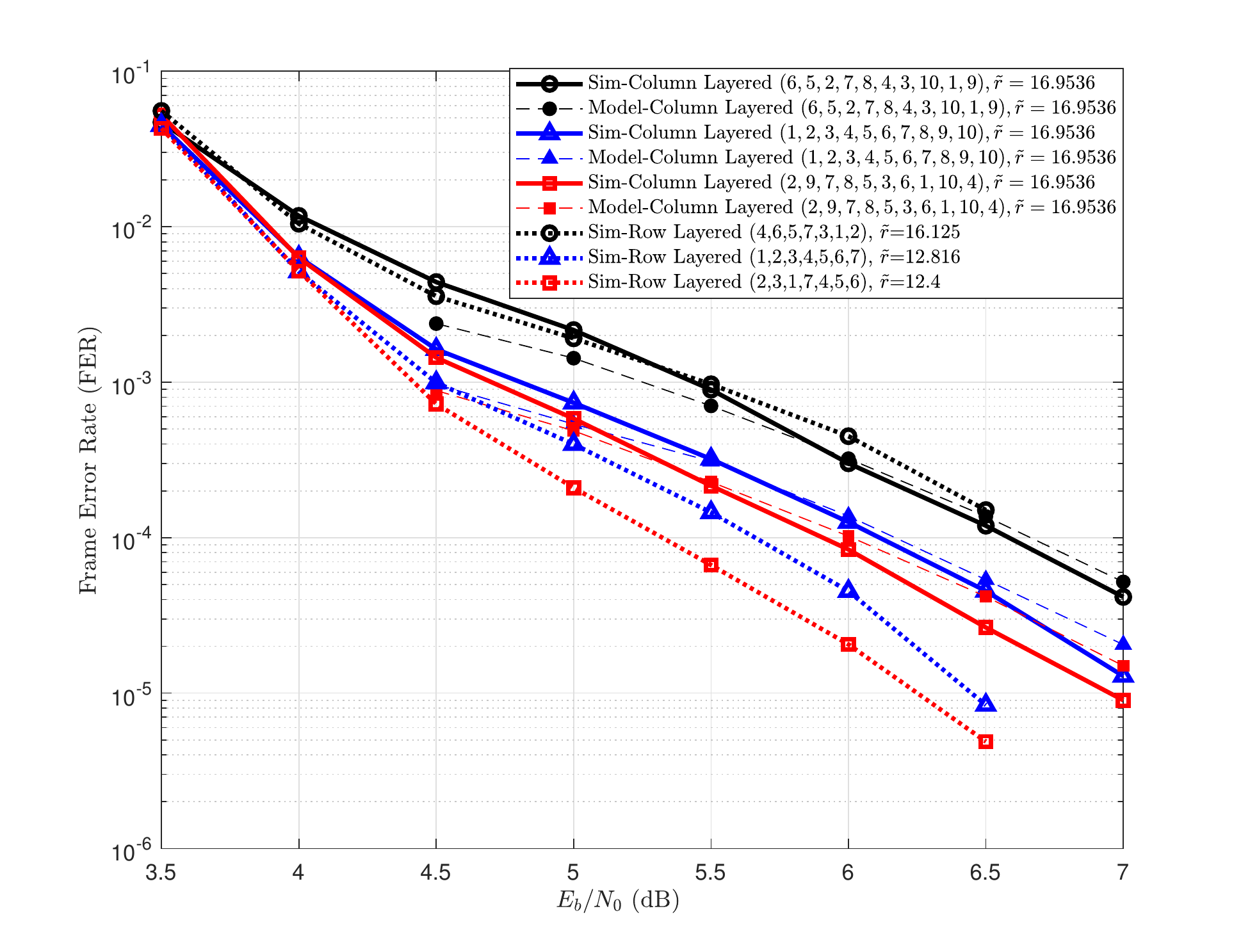}
\vspace{-0.8cm}
\caption{The simulation and estimation results of layered SPA applied to $\mathcal{C}_1$ with different schedules. The saturation level is $15.75$ and the maximum number of iterations is $30$.}
\label{qc640_collay_diffSched_15_75}
\end{figure}

To find a schedule with low error floor, in the next step, we choose a few schedules that resulted in the lowest error rates for the $(5,5)$ LETS in the first step. 
We then apply our estimation technique based on accurate DE to the chosen schedules to find the one that has the lowest error floor.
As a result, we find the schedule corresponding to the permutation $(2,9,7,8,5,3,6,1,10,4)$, shown in Fig. \ref{colOptimization55} with a full square. 
For comparison, we have also selected one of the schedules with the worst error floor, $(6,5,2,7,8,4,3,10,1,9)$, as well as 
the original one, $(1,2,3,4,5,6,7,8,9,10)$. These schedules are specified in Fig. \ref{colOptimization55} by a full circle and a full triangle, respectively. 
The simulation and estimation results of these three schedules are presented in Fig. \ref{qc640_collay_diffSched_15_75}. 
The maximum number of iterations and the saturation level are $I_{max}=30$ and $15.75$, respectively. 
For comparison, we have also presented the FER of different row layered schedules, including the best and the worst schedules, as reported in [my paper], in Fig. \ref{qc640_collay_diffSched_15_75}. 
As can be seen, all the estimation results match with the corresponding simulations rather closely. 
The results show that, unlike the row layered schedules where there is a substantial difference between the error floor of the best and the worst schedules, the difference in the error floor of the best and the worst column layered schedules is rather small (less than an order of magnitude). This can be attributed to the fact that the dominant eigenvalues of transition matrices of various column layered schedules are the same, and thus the difference in the error floor performance is mainly only due to the difference in the distribution of external messages entering the LETS subgraph. 
We also note that the dominant eigenvalue of different column layered schedules, $\tilde{r}=16.9536$, is larger than the maximum 
dominant eigenvalue among all row layered schedules, i.e., $16.125$. One thus expects 
the column layered decoding to perform, on average, worse than the row layered decoding for $\mathcal{C}_1$.
 

We now discuss the error floor of $\mathcal{C}_2$ under column layered SPA. An exhaustive list of $(a,b)$ LETSs of this code within the range $a \leq 8$ and $b \leq 2$ 
are reported in [my paper, Table III]. The dominant class of LETSs for this code is the $(7,1)$ class. This class has $8$ different non-isomorphic structures that are partitioned into $10$ TSLP groups, each of size $24$~[my paper]. 
Due to the large number of layer permutations for this code ($24!$), 
we need to reduce the search space to be able to find a schedule with low error floor. Our experiments show that for the original order of column layers, i.e., $(1, 2, 3, \ldots, 23, 24)$, one of the ten $(7,1)$ TSLP groups, denoted by $(7,1)_{10}$ in [my paper], is the most harmful group. 
The dominant eigenvalue of the transition matrices of the LETSs in this group under the original schedule 
is $\tilde{r}=7.3547$. The number of column layers involved in LETSs of $(7,1)_{10}$ is $J=7$. Therefore, there exist $7!=5040$ 
different permutations of layers within these TSs. These permutations result in transition matrices with $10$ different dominant eigenvalues 
ranging from $6.757$ to $13.877$. Also, there are $854$ different permutations (out of $5040$) whose 
corresponding transition matrices have the minimum dominant eigenvalue of $\tilde{r}=6.757$. 
For each of these $854$ layer permutations of $(7,1)_{10}$, we generate $100$ random column layer permutations of the whole parity check matrix such 
that the corresponding order of layers in the $(7,1)_{10}$ group is preserved. We then examine the resulted $85400$ column layer schedules to find one with the best error floor. The process just described reduces the size of the search space by a factor of $24!/85400 \approx 7.3 \times 10^{18}$. 

\begin{figure}
\centering
\includegraphics[width=3.5in]{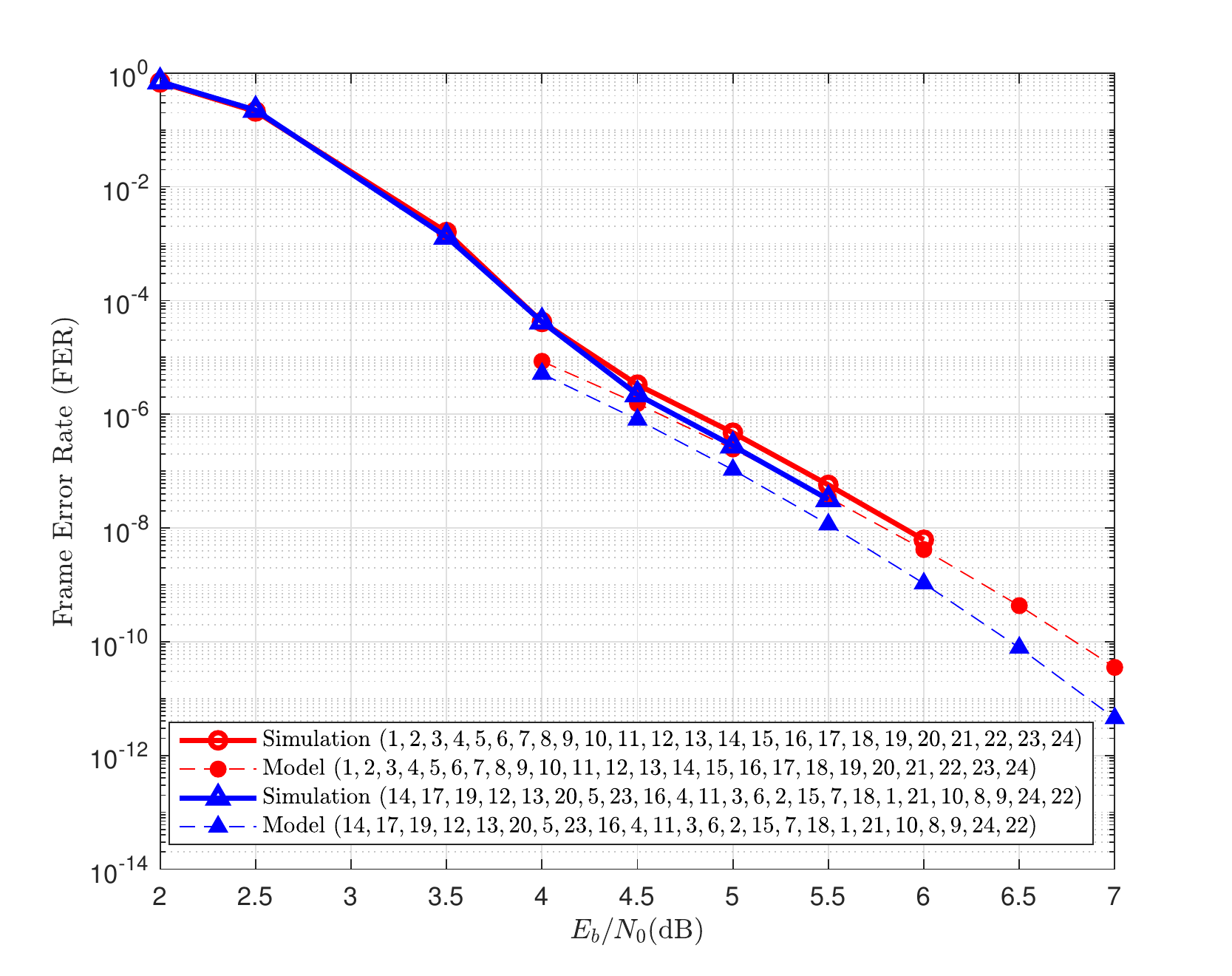}
\vspace{-0.8cm}
\caption{Simulation and estimation results of $\mathcal{C}_2$ under SPA with different column layered schedules. The saturation level and the maximum number of iterations are $15.75$ and $30$, respectively. }
\label{ComparisonWimaxColLay7itS4}
\end{figure}

By using a single application of DE, as discussed in Section \ref{col_opt_Sec}, the total failure rate of $(7,1)$ LETSs for different schedules is approximated. 
Then a handful of schedules with the lowest failure rates are chosen for the examination by the application of accurate DE in the state-space model. 
As a result, we obtain the schedule corresponding to the permutation $(14, 17, 19, 12, 13, 20,$ $ 5, 23, 16, 4, 11, 3, 6, 2, 15, 7, 18, 1, 21, 10, 8, 9, 24, 22)$ as one of the best schedules. The simulation and estimation results of this schedule as well as the schedule corresponding to the original order of column layers are presented in Fig. \ref{ComparisonWimaxColLay7itS4}. The maximum number of iterations and the saturation level are $I_{max}=30$ and $15.75$, respectively. 
It can be seen that the error floor performance of the optimized schedule is slightly better than that of the original column permutations. 
In fact, the original order of column layers, itself, is among good schedules for $\mathcal{C}_2$ in terms of the error floor performance. 
It is worth mentioning that for $\mathcal{C}_2$, the error floor performance of the best row layered schedule~\cite{}, and that of the best column layered schedule found here are practically identical.
    
\vspace{-.3cm}

\section{Conclusion}
In this paper, we developed a linear state-space model for TSs under column layered decoding, also known as shuffled decoding. 
The proposed model was used to estimate the failure rate of TSs and 
to analyze the error floor performance of LDPC codes under saturating SPA with column layered schedule.
We presented simulation results that showed the accuracy of the 
model in estimating the error floor. We also devised a low-complexity search algorithm, based on the proposed model, to find a column layered schedule which minimizes the error floor. We discussed the major differences between the proposed model for column layered decoders and the state-space model of row layered decoders. 
This included the difference in labeling the state variables and the computation of different model parameters. Simulation results for the tested codes also revealed that 
the relative error floor performance of row and column layered decoders depends on the order by which the layers are updated, and either schedule can perform better or worse than the other depending on the updating order of the layers. The same applies to their relative performance with respect to the flooding schedule.\vspace{-.3cm}

\section{Appendix}
Suppose that the systematic form of the flooding schedule transition matrix obtained based on row and column layered schedules are denoted by $\mathbf{A_r}$ and $\mathbf{A_c}$, respectively. (The transition matrix of the flooding schedule changes as a result of the change in the labeling of state variables. The state variable labeling is different in row and column layered schedules.) Then, there exists a permutation matrix $\mathbf{P}_1$ such that
\begin{small}
\begin{equation}\label{P1AcP1T}
\mathbf{A_r}=\mathbf{P}_1\mathbf{A_c}\mathbf{P}_1^T\:,
\end{equation}
\end{small}
where $\mathbf{P}_1^T=\mathbf{P}_1^{-1}$. In \cite[Appendix]{Ali-ITrow}, a symmetric unitary permutation matrix, $\mathbf{P}$, was introduced such that
\begin{small}
\begin{equation}\label{PArTP}
\mathbf{A_r}=\mathbf{P}\mathbf{A_r}^T\mathbf{P}.
\end{equation} 
\end{small}
By substitution of \eqref{P1AcP1T} in \eqref{PArTP}, we have
\begin{small}
\begin{equation}
\mathbf{P}_1\mathbf{A_c}\mathbf{P}_1^T=\mathbf{P}\mathbf{P}_1\mathbf{A_c}^T\mathbf{P}_1^T\mathbf{P}\:,
\end{equation}
\end{small}
or equivalently
\begin{small}
\begin{equation}
\mathbf{A_c}=\mathbf{P}_1^T\mathbf{P}\mathbf{P}_1\mathbf{A_c}^T\mathbf{P}_1^T\mathbf{P}\mathbf{P}_1\:.
\end{equation}
\end{small}
By defining a new unitary permutation matrix $\mathbf{P'}=\mathbf{P}_1^T\mathbf{P}\mathbf{P}_1$, where $\mathbf{P'} \mathbf{P'}=\mathbf{I}$, we can write
\begin{small}
\begin{equation}
\mathbf{A_c}=\mathbf{P'}\mathbf{A_c}^T\mathbf{P'}\:.
\end{equation}
\end{small}
In the proof of Proposition 3 in \cite[Appendix]{Ali-ITrow}, it is sufficient to replace the matrix $\mathbf{P}$ with    $\mathbf{P'}=\mathbf{P}_1^T\mathbf{P}\mathbf{P}_1$, and follow the same steps to prove that the corresponding dominant eigenvalue of the transition matrix in column layered schedule is invariant to the reversing of the layers' order.


\vspace{-1cm}
\begin{thebibliography}{10}
\bibitem{mao2001heuristic}
Y.~Mao and A.~H. Banihashemi, ``A heuristic search for good low-density parity-check codes at short block lengths,'' in {\em Proc. IEEE Int. Conf. Comm., Helsinki, Finland}, pp.~41--44, Jun. 2001.

\bibitem{Tian-2004}
   T. Tian, C. Jones, J. D. Villasenor, and R. D. Wesel, ``Selective avoidance of cycles in irregular {LDPC} code construction,”
   {\em IEEE Trans. Commun.}, vol. 52, pp.~1242--1247, Aug. 2004.

\bibitem{xiao2004improved}
	H.~Xiao and A.~H. Banihashemi, ``Improved progressive-edge-growth ({PEG})
	construction of irregular {LDPC} codes,'' {\em IEEE Commun. Lett.}, vol.~8,
	no.~12, pp.~715--717, Dec. 2004.

\bibitem{Peg}
X.-Y. Hu, E.~Eleftheriou, and D.-M. Arnold, ``Regular and irregular progressive edge-growth tanner graphs,'' {\em IEEE Trans. Inf. Theory}, vol.~51, no.~1, pp.~386--398, Jan. 2005.
\bibitem{Ivkovic}
M. Ivkovic, S. K. Chilappagari, and B. Vasic, ``Eliminating trapping sets in low-density parity-check codes by using Tanner
graph covers,'' {\em IEEE Trans. Inf. Theory}, vol. 54, no. 8, pp. 3763--3768, Aug. 2008.

\bibitem{zheng2010constructing}
	X.~Zheng, F.~C.-M. Lau, and C.~K. Tse, ``Constructing short-length irregular
	{LDPC} codes with low error floor,'' {\em IEEE Trans. Commun.}, vol.~58,
	no.~10, pp.~2823--2834, Oct. 2010.

\bibitem{Asvadi}
R. Asvadi, A. H. Banihashemi, and M. Ahmadian-Attari, ``Lowering the error floor of LDPC codes using cyclic liftings,'' {\em
IEEE Trans. Inf. Theory}, vol. 57, no. 4, pp. 2213--2224, Apr. 2011.

\bibitem{Khaz}
S. Khazraie, R. Asvadi and A. H. Banihashemi, ``A PEG construction of finite-length LDPC codes with low error
floor,'' {\em IEEE Commun. Lett.}, vol. 16, pp. 1288--1291, Aug. 2012.

\bibitem{Nguyen}
D. V. Nguyen, S. K. Chilappagari, M. W. Marcellin, and B. Vasic, ``On the construction of structured LDPC codes free of
small trapping sets,'' {\em IEEE Trans. Inf. Theory}, vol. 58, no. 4, pp. 2280--2302, Apr. 2012.

\bibitem{Tao-2018}
X. Tao, Y. Li, Y. Liu, and Z. Hu, “On the construction of {LDPC} codes free of small trapping sets by controlling cycles,” {\em IEEE Commun. Lett.},
vol. 22, no. 1, pp. 9--12, Jan. 2018.	


\bibitem{Sima-CL1}
S. Naseri and A. H. Banihashemi, ``Construction of girth-8 {QC}-{LDPC} codes free of small trapping sets," {\em IEEE Commun. Lett.}, 
vol. 23, no. 11, pp. 1904--1908, Nov. 2019.

\bibitem{Sima-CL2}
S. Naseri and A. H. Banihashemi, ``Spatially coupled {LDPC} codes with small constraint length and low error floor," {\em IEEE Commun. Lett.}, vol. 24, no. 2, pp. 254--258, Feb. 2020.

\bibitem{Bashir-TCOM}
B. Karimi and A. H. Banihashemi, ``Construction of {QC} {LDPC} codes with low error floor by efficient systematic search and elimination of trapping sets," {\em IEEE Trans. Commun.}, vol. 68, no. 2, pp. 697--712, Feb. 2020.

\bibitem{Bashir-TCOM2}
B. Karimi and A. H. Banihashemi, ``Construction of irregular protograph-based QC-LDPC codes with low error floor," {\em IEEE Trans. Commun.}, vol. 69, no. 1, pp. 3--18, Jan. 2021

\bibitem{Sima-TCOM}
S. Naseri and A. H. Banihashemi, ``Construction of time invariant spatially coupled LDPC codes free of small trapping sets," to appear in {\em IEEE Trans. Commun.}, available on IEEExplore.

\bibitem{danesh}
E. Cavus and B. Daneshrad, ``A performance improvement and error floor avoidance technique for belief propagation
decoding of LDPC codes,'' in {\em Proc. 16th IEEE Int. Symp. Personal, Indoor Mobile Radio Commun.}, Los Angeles, CA,
USA, Sep. 2005, pp. 2386--2390.

\bibitem{Ryan2}
Y. Han and W. E. Ryan, ``LDPC decoder strategies for achieving low error floors,'' in {\em Proc. Inform. Theory Appl. Workshop},
San Diego, CA, USA, Jan. 2008, pp. 277--286.

\bibitem{Ryan1}
Y. Zhang and W. E. Ryan, ``Toward low LDPC-code floors: a case study,'' {\em IEEE Trans. Commun.}, vol. 57, no. 6, pp. 1566--1573, Jun. 2009.

\bibitem{Kyung}
G. B. Kyung and C.-C. Wang, ``Finding the exhaustive list of small fully absorbing sets and designing the corresponding low error-floor decoder,'' {\em IEEE Trans. Commun.}, vol. 60, no. 6, pp. 1487--1498, Jun. 2012.

\bibitem{Zhang1}
S. Zhang and C. Schlegel, ``Controlling the error floor in LDPC decoding,'' {\em IEEE Trans. Commun.}, vol. 61, no. 9, pp. 3566--3575, Sep. 2013.

\bibitem{TB}
S. Tolouei and A. H. Banihashemi, ``Lowering the error floor of LDPC codes using multi-step quantization,'' {\em IEEE Commun. Lett.}, vol. 18, no. 1, pp. 86--89, Jan. 2014.

\bibitem{zhang_quasiuniform}
X. Zhang and P. H. Siegel, ``Quantized iterative message passing decoders with low error floor for LDPC codes,'' {\em IEEE Trans. Commun.}, vol. 62, no. 1, pp. 1--14, January 2014.

\bibitem{TSbreaking}
S. Kang, J. Moon, J. Ha and J. Shin, ``Breaking the trapping sets in LDPC codes: Check node removal and collaborative decoding," {\em IEEE Trans. Commun.}, vol. 64, no. 1, pp. 15--26, Jan. 2016.

\bibitem{homayoon2020}
H. Hatami, D. G. M. Mitchell, D. J. Costello and T. E. Fuja, ``A threshold-based min-sum algorithm to lower the error floors of quantized LDPC decoders,'' {\em IEEE Trans. Commun.}, vol. 68, no. 4, pp. 2005-2015, April 2020.



\bibitem{Wang}
C. C. Wang, S. R. Kulkarni, and H. V. Poor, ``Finding all small error-prone substructures in {LDPC} codes,” {\em IEEE Trans. Inf. Theory}, vol. 55,
no. 5, pp. 1976--1999, May 2009.

\bibitem{mehdi2012}
M.~Karimi and A.~H. Banihashemi, ``Efficient algorithm for finding dominant  trapping sets of {LDPC} codes,'' {\em IEEE Trans. Inf. Theory}, vol.~58, no.~11, pp.~6942--6958, Nov. 2012.

\bibitem{mehdi2014}
M.~Karimi and A.~H. Banihashemi, ``On characterization of elementary trapping sets of variable-regular {LDPC} codes,'' {\em IEEE Trans. Inf. Theory}, vol.~60, no.~9, pp.~5188--5203, Sep 2014.


\bibitem{yoones2015}
Y.~Hashemi and A.~Banihashemi, ``On characterization and efficient exhaustive  search of elementary trapping sets of variable-regular {LDPC} codes,'' {\em IEEE Commun. Lett.}, vol.~19, pp.~323--326, Mar. 2015.

\bibitem{hashemireg}
Y.~Hashemi and A.~H. Banihashemi, ``New characterization and efficient exhaustive search algorithm for leafless elementary trapping sets of variable-regular LDPC codes,'' {\em IEEE Trans. Inf. Theory}, vol. 62, no. 12, pp. 6713--6736, Dec. 2016. 

\bibitem{hashemiireg}
Y.~Hashemi and A.~H. Banihashemi, ``Characterization of elementary trapping sets in irregular LDPC codes and the corresponding efficient exhaustive search algorithms,'' {\em IEEE Trans. Inf. Theory}, vol. 64, no. 5, pp. 3411--3430, May 2018.

\bibitem{richardson}
T.~Richardson, ``Error floors of {LDPC} codes,'' in {\em Proc. 41th annual   Allerton conf. on commun. control and computing}, Monticello, IL, USA, Oct. 2003, pp.~1426--1435.

\bibitem{Sun_phd}
J. Sun, ``Studies on graph--based coding systems,'' Ph.D. dissertation,
Dept. Elect. Eng., Ohio State Univ., Columbus, OH, USA,
2004.

\bibitem{Cole}
C. A. Cole, S. G. Wilson, E. K. Hall, and T. R. Giallorenzi, ``A general
method for finding low error rates of LDPC codes,'' {\em submitted to IEEE
Trans. Inf. Theory}, May 2006.

\bibitem{LaraIS}
L. Dolecek, Z. Zhang, M. Wainwright, V. Anatharam, and B. Nikolic. ``Evaluation of the low frame error rate performance of LDPC codes using importance sampling,'' in {\em Proc. IEEE Inf. Theory Workshop}, Lake Tahoe, CA, Sep. 2–6, 2007, pp. 202--207.

\bibitem{XB-2007}
H. Xiao and A. H. Banihashemi, ``Estimation of bit and frame error rates of finite-length low-density parity-check codes on binary symmetric channels," {\em IEEE Trans. Commun.}, vol. 55, no. 12, pp. 2234--2239, Dec. 2007.

\bibitem{daneshrad}
E. Cavus, C. L. Haymes and B. Daneshrad, ``Low BER performance estimation of LDPC codes via application of importance sampling to trapping sets,'' {\em IEEE Trans. Commun.}, vol. 57, no. 7, pp. 1886--1888, Jul. 2009.

\bibitem{Lara_SP}
L. Dolecek, P. Lee, Z. Zhang, V. Anatharam, B. Nikolic, and M. J. Wainwright,
``Predicting error floors of structured LDPC codes: deterministic
bounds and estimates,'' {\em IEEE J. Sel. Areas Commun.}, vol. 27, no. 6, pp.
908--917, Aug. 2009.

\bibitem{Ontology}
B. Vasić, S. K. Chilappagari, D. V. Nguyen and S. K. Planjery, ``Trapping set ontology,'' in {\em Proc. 47th Allerton Conf.,} Monticello, IL, 2009, pp. 1--7.

\bibitem{Hu_magneticIS}
X. Hu, Z. Li, B. Kumar, and R. Barndt, ``Error floor estimation of long LDPC codes on magnetic recording channels,'' {\em IEEE Trans. Magn.}, vol. 46, no. 6, pp. 1836--1839, Jun. 2010.

\bibitem{Schleg}
C. Schlegel and S. Zhang, ``On the dynamics of the error floor behavior
in (regular) LDPC codes,'' {\em IEEE Trans. Inf. Theory}, vol. 56, no. 7,
pp. 3248--3264, Jul. 2010.

\bibitem{Xiao}
H. Xiao, A. H. Banihashemi, and M. Karimi, ``Error rate estimation of
low-density parity-check codes decoded by quantized soft-decision iterative
algorithms,'' {\em IEEE Trans. Commun.}, vol. 61, no. 2, pp. 474--484,
Feb. 2013.

\bibitem{Sina} S. Tolouei and A. H. Banihashemi, ``Fast and accurate error floor estimation of quantized iterative decoders for variable-regular LDPC codes,'' {\em IEEE Comm. Lett.}, vol. 18, no. 8, pp. 1283--1286, Aug. 2014.

\bibitem{But_SS}
B. K. Butler and P. H. Siegel, ``error floor approximation for LDPC codes in the AWGN channel,'' {\em IEEE Trans. Inf. Theory}, vol. 60, no. 12, pp. 7416--7441, Dec. 2014.



\bibitem{Homayoon_SP}
H. Hatami, D. G. M. Mitchell, D. J. Costello and T. E. Fuja, ``Performance bounds and estimates for quantized LDPC decoders,'' {\em IEEE Trans. Commun.}, vol. 68, no. 2, pp. 683--696, Feb. 2020.

\bibitem{Ali-TCOM}
A. Farsiabi and A. H. Banihashemi, ``Error floor estimation of LDPC decoders - A code independent approach to measuring the harmfulness of trapping sets," {\em IEEE Trans. Commun.}, vol. 68, no. 5, pp. 2667--2679, May 2020.

\bibitem{AS_threshold}
A. Tomasoni, S. Bellini and M. Ferrari, ``Thresholds of absorbing sets in low-density parity-check codes,'' {\em IEEE Trans. Commun.}, vol. 65, no. 8, pp. 3238--3249, Aug. 2017.

\bibitem{Urbank}
T. J. Richardson, M. A. Shokrollahi and R. L. Urbanke, ``Design of capacity-approaching irregular low-density parity-check codes,'' {\em IEEE Trans. Inf. Theory,} vol. 47, no. 2, pp. 619-637, Feb 2001.

\bibitem{Milen}
O. Milenkovic, E. Soljanin, and P. Whiting, ``Asymptotic spectra of trapping sets in regular and irregular LDPC code
ensembles,'' {\em IEEE Trans. Inf. Theory}, vol. 53, no. 1, pp. 39--55, Jan. 2007.

\bibitem{L1}
M. Mansour, N. Shanbhag, ``High-throughput LDPC decoders,'' {\em IEEE Trans. Very Large Scale Integr. (VLSI) Syst.}, vol. 11, no. 6, pp. 976--996, Dec. 2003.


\bibitem{L2}
D. Hocevar, ``A reduced complexity decoder architecture via layered decoding of LDPC codes,'' in {\em Proc. IEEE Workshop Signal Processing and Systems (SIPS.04)}, Austin, TX, Oct. 2004, pp. 107--112.

\bibitem{XiaoSchedule}
H. Xiao and A. H. Banihashemi, ``Graph-based message-passing schedules for decoding LDPC codes," {\em IEEE Trans. Commun.}, vol. 52, no. 12, pp. 2098--2105, Dec. 2004.

\bibitem{NouhSchedule}
A. Nouh and A. H. Banihashemi, ``Reliability-based schedule for bit-flipping decoding of low-density parity-check codes," {\em IEEE Trans. Commun.}, vol. 52, no. 12, pp. 2038--2040, Dec. 2004.

\bibitem{L4}
T. Brack, M. Alles, F. Kienle, N. Wehn, ``A synthesizable IP core for WiMAX 802.16e LDPC code decoding,'' in {\em Proc. IEEE 17th Int. Symp. Personal Indoor and Mobile Radio Communications}, Sept. 2006, pp. 1--5.

\bibitem{L9}
Z. Wang, Z. Cui, ``Low-complexity high-speed decoder design for quasi-cyclic LDPC codes,'' {\em IEEE Trans. VLSI Syst.}, vol. 15, no. 1, pp. 104--114, Jan. 2007.

\bibitem{L6}
K. Gunnam, G. Choi, M. Yeary, M. Atiquzzaman, ``VLSI architectures for layered decoding for irregular LDPC codes of WiMax,'' in {\em Proc. IEEE Int. Conf. Commun. (ICC)}, June 2007, pp. 4542--4547.




\bibitem{L8}
E. Sharon, S. Litsyn, J. Goldberger, ``Efficient serial message-passing schedules for LDPC decoding,'' {\em IEEE Trans. Inf. Theory}, vol. 53, no. 11, pp. 4076--4091, Nov. 2007.



\bibitem{L3}
C.-H. Liu, S.-W. Yen, C.-L. Chen, H.-C. Chang, C.-Y. Lee, Y.-S. Hsu, S.-J. Jou, ``An LDPC decoder chip based on self-routing network for IEEE 802.16e applications,'' {\em IEEE J. Solid-State Circuits}, vol. 43, no. 3, pp. 684--694, March 2008.

\bibitem{L5}
K. Zhang, X. Huang and Z. Wang,
``High-throughput layered decoder implementation for quasi-cyclic LDPC codes,'' {\em IEEE J. Sel. Areas Commun.}, vol. 27, no. 6, pp. 985--994, August 2009.

\bibitem{L7}
Z. Cui, Z. Wang, X. Zhang, ``Reduced-complexity column-layered decoding and implementation for LDPC codes,'' {\em IET Commun.}, vol. 5, no. 15, pp. 2177--2186, 2011.

\bibitem{CL1}
Z. Wang, Z. Cui, and J. Sha, ``VLSI design for low-density parity-check code decoding,'' {\em IEEE Circuits Syst. Mag.}, vol. 11, no. 1, pp. 52–69, Feb. 2011.

\bibitem{CL2}
S. Kumawat et al., ``High-throughput LDPC-decoder architecture using efficient comparison techniques and dynamic multi-frame processing schedule,'' {\em IEEE Trans. Circuits Syst. I, Reg. Papers}, vol. 62, no. 5, pp. 1421–1430, May 2015.

\bibitem{CL3}
J. Zhang and M.-P.-C. Fossorier, ``Shuffled iterative decoding,'' {\em IEEE
Trans. Commun.}, vol. 53, no. 2, pp. 209–213, Feb. 2005.

\bibitem{CL4}
F. Guilloud, E. Boutillon, J. Toush, and J. Danger, ``Generic description and synthesis of LDPC decoders,'' {\em IEEE Trans. Commun.}, vol. 55, no. 11, pp. 2084–2091, Nov. 2007.

\bibitem{CL5}
Z. Cui, Z. Wang, X. Zhang, and Q. Jia, ``Efficient decoder design for
high-throughput LDPC decoding,'' in {\em Proc. IEEE Asia Pacific Conf. on
Circuits and Syst.}, Dec. 2008, pp. 1640–1643.

\bibitem{CL6}
Y. L. Ueng, C. J. Yang, and C. J. Chen, ``A shuffled message-passing
decoding method for memory-based LDPC decoders,'' in {\em Proc. IEEE
ISCAS 2009}, May 2009, pp. 892–895.

\bibitem{CL7}
J. Lin, J. Sha, Z. Wang, and L. Li, ``An improved min-sum based
column-layered decoding algorithm for LDPC codes,'' in {\em Proc. IEEE
Workshop on Signal Processing Systems (SIPS)}, 2009

\bibitem{Ali-ITrow}
A. Farsiabi and A. H. Banihashemi, ``Error floor analysis of LDPC row layered decoders," {\em IEEE Trans. Inf. Theory}, revised in Feb. 2021, available at http://arxiv.org/abs/2104.06867

\bibitem{InfromedDynamic_wesel_2010}
A. I. V. Casado, M. Griot and R. D. Wesel, ``LDPC decoders with informed dynamic scheduling,'' {\em IEEE Trans. Commun.}, vol. 58, no. 12, pp. 3470--3479, December 2010.

\bibitem{M2I2}
H. Lee and Y. Ueng, ``LDPC decoding scheduling for faster convergence and lower error floor," {\em IEEE Trans. Commun.}, vol. 62, no. 9, pp. 3104--3113, Sept. 2014.

\bibitem{vasic_horizontal}
N. Raveendran and B. Vasic, ``Trapping set analysis of horizontal layered decoder," in {\em Proc. Int.
Conf. Commun. (ICC)}, Kansas City, MO, 2018, pp. 1--6.



\bibitem{Angarita_MS_2014}
F. Angarita, J. Valls, V. Almenar and V. Torres, ``Reduced-complexity min-sum algorithm for decoding LDPC codes with low error-floor," {\em IEEE Trans. Circuits Syst. I}, vol. 61, no. 7, pp. 2150--2158, July 2014.

\bibitem{BackTrack_hard_2011}
X. Chen, J. Kang, S. Lin and V. Akella, ``Hardware implementation of a backtracking-based reconfigurable decoder for lowering the error floor of quasi-cyclic LDPC codes," {\em IEEE Trans. Circuits Syst. I}, vol. 58, no. 12, pp. 2931--2943, Dec. 2011.


\bibitem{IDS_kim_2012let}
S. Kim, ``Trapping set error correction through adaptive informed dynamic scheduling decoding of LDPC codes," {\em IEEE Commun. Lett.}, vol. 16, no. 7, pp. 1103--1105, July 2012.











\bibitem{butler_numerical}
B. K. Butler and P. H. Siegel, ``Numerical issues affecting LDPC error floors,''  in {\em Proc. IEEE Global Telecommun. Conf.}, Anaheim, CA, 2012, pp. 3201--3207.

\bibitem{MKarimi_girth}
M. Karimi and A. H. Banihashemi, ``On the girth of quasi cyclic protograph LDPC codes," {\em IEEE Trans. Inf. Theory}, vol. 59, no. 7, pp. 4542-4552, July 2013.

\bibitem{Meyer}
C. D. Meyer, {\em Matrix Analysis and Applied Linear Algebra}. Philadelphia, PA, USA: SIAM, 2000.

\bibitem{varga}
R. S. Varga, {\em Matrix Iterative Analysis}, 2nd ed, Berlin: Springer, 2000.

\bibitem{noutsos_slide}
D. Noutsos, ``Perron-Frobenius theory and some extensions", Como,
Italy, May 2008, [Presentation Slides]. Available: http://www.math.uoi.
gr/ dnoutsos/Papers-pdf-files/slide-perron.pdf.

\bibitem{horn}
R. A. Horn, C. R. Johnson, {\em Matrix Analysis}, Cambridge, U.K.: Cambridge Univ. Press, 1985.

\bibitem{wimax_standard}
{\em IEEE Standard for Local and Metropolitan Area Networks—Part 16:
Air Interface for Fixed and Mobile Broadband Wireless Access Systems
Amendment 2: Physical and Medium Access Control Layers for Combined
Fixed and Mobile Operation in Licensed Bands and Corrigendum
1}, IEEE Standard 802.16e-2005 and 802.16-2004/Cor 1-2005, Feb. 2006.
\end{thebibliography}
\end{document}